\documentclass[amsmath,showpacs,nofootinbib,12pt]{revtex4-2}
\usepackage{graphicx}
\usepackage{dcolumn}
\usepackage{bm}
\usepackage{color} 
\usepackage{slashed}
\usepackage{amsfonts}
\begin{document}
\newcommand{\hs}{\hspace*{0.5cm}}
\newcommand{\vs}{\vspace*{0.5cm}}
\newcommand{\be}{\begin{equation}}
\newcommand{\ee}{\end{equation}}
\newcommand{\bea}{\begin{eqnarray}}
\newcommand{\eea}{\end{eqnarray}}
\newcommand{\ben}{\begin{enumerate}}
\newcommand{\een}{\end{enumerate}}
\newcommand{\bde}{\begin{widetext}}
\newcommand{\ede}{\end{widetext}}
\newcommand{\nn}{\nonumber}
\newcommand{\crn}{\nonumber \\}
\newcommand{\Tr}{\mathrm{Tr}}
\newcommand{\non}{\nonumber}
\newcommand{\noi}{\noindent}
\newcommand{\al}{\alpha}
\newcommand{\la}{\lambda}
\newcommand{\bet}{\beta}
\newcommand{\ga}{\gamma}
\newcommand{\va}{\varphi}
\newcommand{\om}{\omega}
\newcommand{\pa}{\partial}
\newcommand{\+}{\dagger}
\newcommand{\fr}{\frac}
\newcommand{\bc}{\begin{center}}
\newcommand{\ec}{\end{center}}
\newcommand{\Ga}{\Gamma}
\newcommand{\de}{\delta}
\newcommand{\De}{\Delta}
\newcommand{\ep}{\epsilon}
\newcommand{\varep}{\varepsilon}
\newcommand{\ka}{\kappa}
\newcommand{\La}{\Lambda}
\newcommand{\si}{\sigma}
\newcommand{\Si}{\Sigma}
\newcommand{\ta}{\tau}
\newcommand{\up}{\upsilon}
\newcommand{\Up}{\Upsilon}
\newcommand{\ze}{\zeta}
\newcommand{\ps}{\psi}
\newcommand{\Ps}{\Psi}
\newcommand{\ph}{\phi}
\newcommand{\vph}{\varphi}
\newcommand{\Ph}{\Phi}
\newcommand{\Om}{\Omega}
\newcommand{\AdrHEPC}{Phenikaa Institute for Advanced Study and Faculty of Basic Science, Phenikaa University, Yen Nghia, Ha Dong, Hanoi 100000, Vietnam}

\title{Phenomenology of a minimal extension of the standard model with a family-dependent gauge symmetry} 

\author{Duong Van Loi}
\email{loi.duongvan@phenikaa-uni.edu.vn (corresponding author)}
\author{Cao H. Nam}
\email{nam.caohoang@phenikaa-uni.edu.vn}
\author{Phung Van Dong} 
\email{dong.phungvan@phenikaa-uni.edu.vn}

\affiliation{\AdrHEPC} 
\date{\today}

\begin{abstract}
We consider a gauge symmetry extension of the standard model given by $SU(3)_C\otimes SU(2)_L\otimes U(1)_X\otimes U(1)_N\otimes\mathbb{Z}_2$ with minimal particle content, where $X$ and $N$ are family dependent but determining the hypercharge as $Y=X+N$, while $\mathbb{Z}_2$ is an exact discrete symmetry. In our scenario, $X$ (while $N$ is followed by $X-Y$) and $\mathbb{Z}_2$ charge assignments are inspired by the number of fermion families and the stability of dark matter, as observed, respectively. We examine the mass spectra of fermions, scalars, and gauge bosons, as well as their interactions, in presence of a kinetic mixing term between $U(1)_{X,N}$ gauge fields. We discuss in detail the phenomenology of the new gauge boson and the right-handed neutrino dark matter stabilized by $\mathbb{Z}_2$ conservation. We obtain parameter spaces simultaneously satisfying the recent CDF $W$-boson mass, electroweak precision measurements, particle colliders, as well as dark matter observables, if the kinetic mixing parameter is not necessarily small.
\end{abstract}

\maketitle

\section{Introduction}

The standard model (SM) \cite{ParticleDataGroup:2022pth} of fundamental particles and interactions predicts, in principle, an arbitrary number of fermion families, which is opposite to only three families observed in the nature. Additionally, the SM fails to explain why neutrinos have small masses and flavor mixing, as it both conserves the lepton number and possesses no right-handed neutrinos. Furthermore, the SM does not contain any candidate for dark matter, which makes up most of the mass of galaxies and galaxy clusters. 

Among attempts to solve the above questions, the 3-3-1 model \cite{Pisano:1992bxx,Frampton:1992wt,Foot:1992rh,Valle:1983dk,Montero:1992jk,Foot:1994ym} is very attractive because it not only addresses the issues of neutrino mass generation \cite{Tully:2000kk,Dias:2005yh,Chang:2006aa,Dong:2006mt,Dong:2008sw,Dong:2010gk,Dong:2010zu,Dong:2011vb,Boucenna:2014ela,Boucenna:2014dia,Boucenna:2015zwa,Okada:2015bxa,Pires:2014xsa,Dias:2010vt,Huong:2016kpa,Reig:2016tuk} and dark matter stability \cite{Fregolente:2002nx,Hoang:2003vj,Filippi:2005mt,deS.Pires:2007gi,Mizukoshi:2010ky,Alvares:2012qv,Profumo:2013sca,Kelso:2013nwa,daSilva:2014qba,Dong:2013ioa,Dong:2014esa,Dong:2015rka,Dong:2013wca,Dong:2014wsa,Dong:2015yra,Huong:2016ybt,Alves:2016fqe,Ferreira:2015wja,Dong:2017zxo}, but also provides a partial answer to the problem of family number, which matches that of colors by anomaly cancelation requirement (cf. \cite{Frampton:1992wt}). Because the 3-3-1 model arranges families nonuniversally, it is hard to be embedded in a GUT which is family universal, by contrast \cite{Huong:2016ybt}. The most minimal unification that contains the 3-3-1 model is a flipped $SU(6)$ or a flipped trinification \cite{Huong:2016kpa}, which leaves a family-dependent abelian factor as flipped, basically being in the same situation with the 3-3-1 model. Obviously, the flipped abelian factor may originate from a more fundamental theory, but this final theory has not emerged yet.\footnote{Gauged family symmetries (see, e.g., \cite{Berezhiani:1990wn}) or string models with different constructions of distinct families (see, for a review, \cite{Langacker:2000ju}) might be a good approach put forward.} Additionally, we do not ascertain which 3-3-1 model and/or its variants are potential intermediate physical phases over the relevant energy scale. Hence, it is worthwhile specifying their simplest remnant---the family-dependent abelian charge---as well as the way this remnant affecting the low energy physics. 

In recent work, we have shown that such an abelian remnant is viable \cite{VanDong:2022cin}. Indeed, we extend $U(1)_Y$ to $U(1)_X\otimes U(1)_N$ such that $X$ is a family-dependent neutral charge furnished by the 3-3-1 model or its variants, while $N$ is necessarily introduced for determining the hypercharge $Y=X+N$ after $X,N$ breaking, thus $N=Y-X$. If we choose $X\sim T_{8L}+T_{8R}$ as of trinification, the most minimal fermion content is revealed. Since known fermion multiplets arise from trinification triplets/antitriplets, we assign $X_{l_L}=x$ for $N_f$ lepton doublets $l_L = (\nu_L,e_L)$, while $X_{q_L}=x$ for $n$ quark doublets and $X_{q_L}=-x$ for the remaining $m=N_f-n$ quark doublets, with $q_L=(u_L,d_L)$. The anomaly $[SU(2)_L]^2 U(1)_X\sim 3X_{q_L}+X_{l_L}$ is canceled over $N_f$ families, if $N_f x+3nx-3mx=0$, thus $N_f=3(m-n)$ to be a multiple of color number, 3. Additionally, since quarks carry different $X$-charge, the flavor-changing neutral currents (FCNCs) appear at the tree-level. The model contains three right-handed neutrinos by anomaly cancelation requirement, which induces small neutrino masses via the seesaw mechanism. 

The $U(1)_X\otimes U(1)_N$ model requires at least three Higgs doublets to generate appropriate quark masses at tree-level. In this work, we consider a minimal Higgs sector only with the standard model Higgs doublet, opposite to \cite{VanDong:2022cin}, besides a scalar singlet for breaking $U(1)_X\otimes U(1)_N\rightarrow U(1)_Y$, as usual. We signify an exact $\mathbb{Z}_2$ symmetry which singles out a right-handed neutrino to be odd, whereas all the other particles, including the remaining two of three right-handed neutrinos, are assigned to be even. It implies that small neutrino masses are generated via the so-called minimal seesaw mechanism only with two right-handed neutrinos \cite{King:1999mb,Frampton:2002qc}. Additionally, the right-handed neutrino that is odd is stabilized by $\mathbb{Z}_2$, providing a dark matter candidate for the model by itself. This type of dark matter scheme is very simple and concise, being studied extensively in the literature \cite{Okada:2010wd,Okada:2012sg,Basak:2013cga,Okada:2016gsh,Okada:2016tci,Cox:2017rgn,Escudero:2018fwn,Nam:2020byw}. In addition, it is interesting that the present model contains two $U(1)$ groups, yielding a kinetic mixing term between two relevant gauge bosons in the Lagrangian. We will prove that the kinetic mixing effect is necessary for this setup properly working. We will find viable parameter spaces which not only explain the $W$-boson mass deviation recently reported by the CDF collaboration \cite{CDF:2022hxs}, but also satisfy electroweak precision measurements, particle colliders, as well as dark matter observables.

The rest of this work is organized as follows. In Sect. \ref{model}, we revisit the $U(1)_X\otimes U(1)_N$ model, interpreting its gauge symmetry, discrete symmetry, particle content, and charge assignment. In Sect. \ref{fermion}, we examine the mass spectrum of fermions and specially that of neutrinos. In Sect. \ref{gaugescalar}, the mass spectra of scalars and gauge bosons are presented. In Sect. \ref{fgint}, we determine the couplings of physical gauge fields with fermions in presence of the kinetic mixing term between two $U(1)$ gauge bosons. In Sect \ref{constraint}, we discuss in detail the experimental constraints on the new physics scale through the recent CDF $W$-boson mass, electroweak precision measurements, and particle colliders. Sect. \ref{DM} is devoted to direct and indirect detections of the right-handed neutrino dark matter. Finally, we summarize our results and make conclusions in Sect. \ref{conclusion}.

\section{\label{model}The model}

The symmetry of the present model is given by
\be SU(3)_C\otimes SU(2)_L\otimes U(1)_X\otimes U(1)_N\otimes \mathbb{Z}_2, \label{symmetry}\ee 
where the first and second factors are the ordinary color charge and weak isospin symmetries, respectively. The factor $\mathbb{Z}_2$ is an exact discrete symmetry, which singles out a right-handed neutrino to be odd. The family-dependent abelian charges $X$ and $N$ would determine the hypercharge after breaking at high energy,  
\be Y=X+N,\label{YXN}\ee 
which is also called a flipped SM \cite{VanDong:2020cjf,Nam:2020byw}. 

Taking $X\sim T_{8L}+T_{8R}$, while $N$ is followed by $N=Y-X$, the field content and charge assignment for each field under symmetry (\ref{symmetry}) are listed in Table \ref{tab1}, which differs from \cite{VanDong:2022cin} in scalar sector and $\mathbb{Z}_2$. Besides the SM particles, the model contains three right-handed neutrino $\nu_{aR}$ necessarily for canceling anomalies and a scalar singlet $\chi$ required for breaking $U(1)_X\otimes U(1)_N\to U(1)_Y$ as well as generating the right-handed neutrino mass scale via $\nu_R\nu_R\chi$ coupling. Hence, we require that $\chi$ possesses a vacuum expectation value (VEV),
\be \langle\chi\rangle=\La/\sqrt2,\ee
satisfying $\La\gg v=246$ GeV for consistency with the SM, where $v$ is the SM Higgs VEV,
\be \langle\phi\rangle = \begin{pmatrix}
0\\
v/\sqrt2
\end{pmatrix}.\ee

As shown in \cite{VanDong:2022cin}, the $X$ charge assignment is inspired by the observed number of fermion families, while the assignment for $N$ charge is related to $X$ through (\ref{YXN}). Such $X$ charge can originate from a trinification \cite{trinification}, $E_6$ \cite{Gursey:1975ki}, or their variants \cite{Singer:1980sw,Valle:1983dk,Pisano:1992bxx,Huong:2016kpa,Dong:2017zxo}. However, a minimal grand unification such as $SU(5)$ \cite{Georgi:1974sy}, $SO(10)$ \cite{Fritzsch:1974nn}, and Pati–Salam model \cite{Pati:1974yy} do not contain this kind of charge. It is verified that all anomalies, such as $[SU(3)_C]^2U(1)_X$, $[SU(3)_C]^2U(1)_N$, $[SU(2)_L]^2U(1)_X$, $[SU(2)_L]^2U(1)_N$, $[\mathrm{gravity}]^2U(1)_X$, $[\mathrm{gravity}]^2U(1)_N$, $[U(1)_X]^2U(1)_N$, $[U(1)_N]^2U(1)_X$, $[U(1)_X]^3$, and $[U(1)_N]^3$, are canceled within the fermion content in Table \ref{tab1}, independent of $x$.
\begin{table}
\bc
\begin{tabular}{lccccc}
\hline\hline 
Fields & $SU(3)_C$ & $SU(2)_L$ & $U(1)_X$ & $U(1)_N$ & $\mathbb{Z}_2$\\
\hline 
$L_{aL}=\begin{pmatrix}
\nu_{aL}\\
e_{aL}\end{pmatrix}$ & $\bf {1}$ & $\bf {2}$ & $x$ & $-1/2-x$ & $+$\\
$e_{aR}$ & $\bf {1}$ & $\bf {1}$ & $x$ & $-1-x$ & $+$\\
$Q_{\al L}=\begin{pmatrix}
u_{\al L}\\
d_{\al L}\end{pmatrix}$ & $\bf {3}$ & $\bf {2}$ & $-x$ & $1/6+x$ & $+$\\
$u_{\al R}$ & $\bf {3}$ & $\bf {1}$ & $-x$ & $2/3+x$ & $+$\\
$d_{\al R}$ & $\bf {3}$ & $\bf {1}$ & $-x$ & $-1/3+x$ & $+$\\
$Q_{3 L}=\begin{pmatrix}
u_{3 L}\\
d_{3 L}\end{pmatrix}$ & $\bf {3}$ & $\bf {2}$ & $x$ & $1/6-x$ & $+$\\
$u_{3 R}$ & $\bf {3}$ & $\bf {1}$ & $x$ & $2/3-x$ & $+$\\ 
$d_{3R}$ & $\bf {3}$ & $\bf {1}$ & $x$ & $-1/3-x$ & $+$\\
$\phi=\begin{pmatrix}
\phi^+_1\\
\phi^0_2
\end{pmatrix}$ & $\bf {1}$ & $\bf {2}$ & 0 & $1/2$ & $+$\\
$\nu_{\al R}$ & $\bf {1}$ & $\bf {1}$ & $x$ & $-x$ & $+$\\
$\nu_{3R}$ & $\bf {1}$ & $\bf {1}$ & $x$ & $-x$ & $-$\\
$\chi$ & $\bf {1}$ & $\bf {1}$ & $-2x$ & $2x$ & $+$\\
\hline\hline
\end{tabular}
\caption[]{\label{tab1}Matter content and charge assignment under the symmetry (\ref{symmetry}), where $a=1,2,3$ and $\al=1,2$ denote family indices, while the parameter $x$ is arbitrarily nonzero.}
\ec
\end{table} 

Concerning $\mathbb{Z}_2$, we interpret it as a residual symmetry arising from a high-scale symmetry breaking of a certain $U(1)$ group, such that under this $U(1)$, the $U(1)$ charge assignment for the first two right-handed neutrinos is different to that for the third right-handed neutrino, but still satisfying the cancellation of all anomalies. (For instance, an allowed charge assignment under $U(1)_{B-L}$ is $[B-L]({\nu_{\al R}})=-4$ and $[B-L]({\nu_{3R}})=5$ \cite{Montero:2007cd,Montero:2011jk,Sanchez-Vega:2014rka,Costa:2019zzy}.) Then, with a consistent $U(1)$ charge assignment, it may lead to that the SM particles and the first two right-handed neutrinos are even, whereas the third right-handed neutrino is odd, under $\mathbb{Z}_2$. It is interesting that the first two right-handed neutrinos are just enough to generate observed light neutrino masses via the minimal seesaw mechanism, with prediction of one massless eigenstate, while the third right-handed neutrino is stable and cannot decay to normal
fields, providing a dark matter candidate.

The total Lagrangian is given by
\be \mathcal{L} = \mathcal{L}_{\mathrm{kinetic}} + \mathcal{L}_{\mathrm{Yukawa}} -V. \ee
The first part contains kinetic terms and gauge interactions,
\bea \mathcal{L}_{\mathrm{kinetic}} &=& \sum_F\bar{F}i\gamma^\mu D_\mu F+\sum_S(D^\mu S)^\dag (D_\mu S)\crn
&&-\fr 1 4 G_{p\mu\nu}G_p^{\mu\nu}-\fr 1 4 A_{j\mu\nu}A_j^{\mu\nu}-\fr 1 4 B_{\mu\nu}B^{\mu\nu}-\fr 1 4 C_{\mu\nu}C^{\mu\nu}-\fr \delta 2 B_{\mu\nu}C^{\mu\nu},\label{kinetic} \eea
where $F$ and $S$ run over fermion and scalar multiplets, respectively. The covariant derivative and field strength tensors are given by
\bea D_\mu &=& \partial_\mu + ig_st_pG_{p\mu} + igT_jA_{j\mu} + ig_X XB_\mu + ig_N N C_\mu,\\
 G_{p\mu\nu} &=& \partial_\mu G_{p\nu}-\partial_\nu G_{p\mu} - g_s f_{pqr}G_{q\mu}G_{r\nu},\\
 A_{j\mu\nu} &=& \partial_\mu A_{j\nu}-\partial_\nu A_{j\mu} - g \epsilon_{jkl}A_{k\mu}A_{l\nu},\\
 B_{\mu\nu} &=& \partial_\mu B_\nu-\partial_\nu B_\mu,\hs C_{\mu\nu} = \partial_\mu C_\nu-\partial_\nu C_\mu, \eea
 where $(g_s,g,g_X,g_N)$, $(t_p,T_j,X,N)$, and $(G_{p\mu},A_{j\mu},B_\mu,C_\mu)$ correspond to coupling constants, generators, and gauge bosons of $(SU(3)_C, SU(2)_L, U(1)_X, U(1)_N)$ groups, respectively. $f_{pqr}$ and $\epsilon_{jkl}$ are the structure constants of $SU(3)_C$ and $SU(2)_L$ groups, respectively. $\delta$ is a kinetic mixing parameter between the two $U(1)$ gauge bosons, $B_\mu$ and $C_\mu$. 
 
Let us rewrite the kinetic terms of $B_\mu$ and $C_\mu$ in the canonical form as $-\fr 1 4 \hat{B}_{\mu\nu}\hat{B}^{\mu\nu}-\fr 1 4 \hat{C}_{\mu\nu}\hat{C}^{\mu\nu}$ by assigning $\hat{B}_{\mu\nu}=B_{\mu\nu}+\delta C_{\mu\nu}$ and $\hat{C}_{\mu\nu} = \sqrt{1-\delta^2}C_{\mu\nu}$. Then, the kinetic energy term of $\hat{C}_\mu$ requires $1-\delta^2>0$ or equivalently $|\delta|<1$ in order for
definitely positive kinetic energy. Additionally, if $SU(2)_L\otimes U(1)_X$ are unified into a larger gauge group, such as 3-3-1 or trinification, $\delta$ is tiny. The reason is that the last term in Eq. (\ref{kinetic}) is manifestly prevented at high energy by gauge invariance, while at low energy, it is generated only at loop level, or through effective interactions \cite{VanDong:2023bpg}. Alternatively, if $U(1)_X$ is independent, as it depends on families whereas more-fundamental theories such GUTs are often family universal, the last term in Eq. ({\ref{kinetic}}) is always allowed by gauge invariance, such constraint on $\delta$ is relaxed. Since the final theory has not determined yet, in this work we consider a general condition for $\delta$, such as $-1<\delta<1$.
 
 The remaining parts of the Lagrangian correspond to the Yukawa Lagrangian,
 \bea
 \mathcal{L}_{\mathrm{Yukawa}} &=&  h^e_{ab}\bar{L}_{aL} \phi e_{bR}+h^\nu_{a\al} \bar{L}_{aL} \tilde{\phi} \nu_{\al R}+ \fr 1 2 f^\nu_a \bar{\nu}^C_{aR} \nu_{aR}\chi\crn
&&+ h^d_{\al \beta} \bar{Q}_{\al L} \phi d_{\beta R}+ h^u_{\al \beta} \bar{Q}_{\al L} \tilde{\phi} u_{\beta R} +h^d_{33}\bar{Q}_{3L} \phi d_{3R}+ h^u_{33} \bar{Q}_{3L} \tilde{\phi} u_{3R}\crn
&&+\fr{h^d_{\al 3}}{M} \bar{Q}_{\al L} \phi\chi d_{3 R}+ \fr{h^u_{\al 3}}{M} \bar{Q}_{\al L} \tilde{\phi}\chi u_{3 R} + \fr{h^d_{3\beta}}{M}\bar{Q}_{3L} \phi\chi^* d_{\beta R}+ \fr{h^u_{3\beta}}{M} \bar{Q}_{3L} \tilde{\phi}\chi^* u_{\beta R}\crn
&&+\mathrm{H.c.},\label{yukawa}\eea and the scalar potential,
\bea 
 V = \mu_1^2\phi^\dag \phi + \mu_2^2\chi^\dag\chi + \la_1(\phi^\dag \phi)^2 + \la_2(\chi^\dag \chi)^2 + \la_3(\phi^\dag \phi)(\chi^\dag \chi).\label{potential}
 \eea
We have assumed $f^\nu$ to be flavor diagonal, without loss of generality. Additionally, $h$'s, $f^\nu$, and $\la$'s are dimensionless, whereas $\mu_{1,2}$ have a mass dimension. We have defined $\tilde{\phi}=i\sigma_2 \phi^*$, where $\sigma_2$ is the second Pauli matrix, and $M$ to be a new physics or cutoff scale, which defines the effective interactions. 

\section{\label{fermion}Fermion mass}

After spontaneous symmetry breaking, fermions obtain a mass through the Yukawa Lagrangian in Eq. (\ref{yukawa}). Concerning charged leptons, their mass matrix is given by
\be [M_e]_{ab}=-h^e_{ab}\frac{v}{\sqrt2}, \ee governed by the $h^e$ coupling. After diagonalizing this matrix, we get the realistic masses of electron, muon, and tau to be $m_e$, $m_\mu$, and $m_\tau$, respectively, and the gauge states $e=(e_1,e_2,e_3)$ and mass eigenstates $e'=(e,\mu,\tau)$ are related by $e_{L,R} = V_{eL,R}e'_{L,R}$, such that 
\be \text{diag}(m_e,m_\mu,m_\tau)=V_{eL}^\dag M_e V_{eR}, \ee
where $V_{eL,R}$ are unitary matrices.

Concerning quarks, we obtain two $3\times 3$ mass matrices, $M_d$ and $M_u$, corresponding to down-type quarks and up-type quarks, respectively. The elements of $M_d$ are
\be [M_d]_{\al\beta} = -h^d_{\al\beta}\frac{v}{\sqrt2}, \hs [M_d]_{33}=-h^d_{33}\frac{v}{\sqrt2}, \ee
\be [M_d]_{\al 3} = -h^d_{\al 3}\fr{v\La}{2M}, \hs [M_d]_{3\beta}=-h^d_{3\beta}\frac{v\La}{2M}.  \ee
The elements of $M_u$ are
\be [M_u]_{\al\beta} = -h^u_{\al\beta}\frac{v}{\sqrt2}, \hs [M_u]_{33}=-h^u_{33}\frac{v}{\sqrt2}, \ee
\be [M_u]_{\al 3} = -h^u_{\al 3}\frac{v\La}{2M}, \hs [M_u]_{3\beta}=-h^u_{3\beta}\frac{v\La}{2M}.  \ee
The small mixing between the third quark family and the first two quark families in CKM matrix can be understood by either $h_{\al 3},h_{3\beta}<h_{\al\beta},h_{33}$ or $M> \La$.  Of course, we can diagonalize the mass matrices $M_{d,u}$ to get the realistic masses of quarks, such as 
\bea \text{diag}(m_d,m_s,m_b)&=&V_{dL}^\dag M_d V_{dR},\\
\text{diag}(m_u,m_c,m_t)&=&V_{uL}^\dag M_u V_{uR},\eea where
the mass eigenstates, $d'=(d,s,b)$ and $u'=(u,c,t)$, are related to the gauge states, $d=(d_1,d_2,d_3)$ and $u=(u_1,u_2,u_3)$, by the unitary matrices, $V_{dL,R}$ and $V_{uL,R}$, such as 
\be d_{L,R} = V_{dL,R}d'_{L,R},\hs u_{L,R} = V_{uL,R}u'_{L,R}, \ee respectively. Notice that the Cabibbo-Kobayashi-Maskawa (CKM) matrix is $V=V^\dagger_{uL}V_{dL}$.

It is important to notice that the third right-handed neutrino $\nu_{3R}$ does not mix with other fields due to the $\mathbb{Z}_2$ conservation. Therefore, the field $\nu_{3R}$ is a physical field by itself, with a mass proportional to $f_3^\nu$, i.e., $m_{\nu_{3R}}=-\La f_3^\nu/\sqrt2$.  

For the remaining neutrinos, $\nu_{aL}$ and $\nu_{\al R}$, their mass generation Lagrangian can be given in form of
\be \mathcal{L}_Y^\nu \supset -\fr 1 2 \begin{pmatrix}\bar{\nu}^C_L & \bar{\nu}_R\end{pmatrix}
\begin{pmatrix}
0 & M_D\\
M^T_D & M_M\end{pmatrix}
\begin{pmatrix}
\nu_L\\
\nu^C_R\end{pmatrix} +\mathrm{H.c.}, \ee
where
\be [M_D]_{a\al}=-\fr{v}{\sqrt2}\left(\begin{array}{cc}h^\nu_{11}&h^\nu_{12}\\ h^\nu_{21}&h^\nu_{22}\\h^\nu_{31}&h^\nu_{32}\end{array}\right), \hs [M_M]_{\al\bet}=-\fr{\La}{\sqrt2}\text{diag}\left(f^\nu_{1},f^\nu_{2}\right),\ee
which are Dirac and Majorana mass matrices, respectively. Hence, the canonical seesaw is automatically implemented due to $\La\gg v$, yielding small masses for observed neutrinos $\nu'_{L}$ as well as large masses for heavy neutrinos $\nu'_{R}$, such as
\bea \text{diag}(m_{\nu'_{1L}},m_{\nu'_{2L}},m_{\nu'_{3L}})&\simeq& -U^TM_DM_M^{-1}M_D^TU=U^Th^\nu (f^\nu)^{-1}(h^\nu)^TU\fr{v^2}{\sqrt2 \La},\\
\text{diag}(m_{\nu'_{1R}},m_{\nu'_{2R}})&\simeq& M_M.\eea Here,
the gauge states are related to the mass eigenstates by
\be \begin{pmatrix}
\nu_L\\ \nu^C_R
\end{pmatrix}\simeq\begin{pmatrix}
1 & \epsilon^*\\
-\epsilon^T & 1
\end{pmatrix}\begin{pmatrix}
U & 0\\
0 & 1
\end{pmatrix}\begin{pmatrix}
\nu_L^\prime\\ \nu^{\prime C}_R
\end{pmatrix},\ee
where $U$ is the Pontecorvo-Maki-Nakagawa-Sakata (PMNS) matrix, given that the charged
leptons are flavor diagonal, i.e. $V_{eL}=1$. Because of $\det (M_DM_M^{-1}M_D^T) = 0$, the present model predicts one massless neutrino eigenstate, while the masses of two remaining neutrinos are completely fixed by observed neutrino mass-squared differences, which is still consistent with the neutrino oscillation data \cite{King:1999mb,Frampton:2002qc}. Since $\nu_L$-$\nu_R$ mixing is quite small, i.e. $\epsilon=M_DM^{-1}_M=h^\nu(f^\nu)^{-1}v/\La\ll 1$, we can approximate $\nu_{aL}\simeq U_{ai}\nu'_{iL}$ and $\nu_{\al R}\simeq \nu'_{\al R}$. 

Using the experimental data $m_{\nu'_L}\sim 0.1$ eV as well as taking $v=246$ GeV, we get
\be \La\sim [(h^\nu)^2/f^\nu]\times 10^{14}\text{ GeV}.  \ee
When $(h^\nu)^2/f^\nu$ is sufficiently small, i.e. $f^\nu\sim 1$ and $h^\nu\sim 10^{-5}$ similar to electron Yukawa coupling, the model predicts a seesaw scale $\La$ in TeV regime. Otherwise, if $(h^\nu)^2/f^\nu\sim 1$, the seesaw scale is in order $\mathcal{O}(10^{14})$ GeV. The low-scale seesaw scenario at TeV is very attractive due to a significant modification for the SM phenomenology as well as a potential discovery of the new physics at LHC. In the present work, we consider only this case. 

We will omit the prime mark in the notation for physical lepton and quark fields, without confusion, since they are subscripted by an index, say $f_{i/j}$, distinct from family index $a/b$. 

\section{\label{gaugescalar}Boson sector}

In what follows, we diagonalize the scalar and gauge boson sectors in order to obtain relevant physical mass spectra. 

\subsection{Scalar sector}

Firstly, we consider the scalar potential in Eq. (\ref{potential}). The necessary conditions for this potential to be bounded from below as well as to yield a desirable vacuum structure are
\be \la_{1,2}>0,\hs \la_3>-2\sqrt{\la_1\la_2},\hs \mu_{1,2}^2<0,\hs |\mu_1|\ll |\mu_2|. \ee

To obtain the potential minimum and physical scalar spectrum, we expand the scalar fields around their VEVs as follows:
\be \phi=\begin{pmatrix}
\phi^+_1\\
\frac{1}{\sqrt2}(v+S_1+iA_1)
\end{pmatrix} , \hs \chi = \frac{1}{\sqrt2}(\La +S_2+iA_2). \ee
Substituting these fields into the scalar potential in Eq. (\ref{potential}), we derive conditions for the potential minimum, such as
\be \La^2=\frac{2(\la_3\mu_1^2-2\la_1\mu_2^2)}{4\la_1\la_2-\la_3^2},\hs v^2=\frac{2(\la_3\mu_2^2-2\la_2\mu_1^2)}{4\la_1\la_2-\la_3^2}. \ee

Using the potential minimum conditions, we obtain physical scalar fields,
\be \phi=\begin{pmatrix}
G^+_W\\
\frac{1}{\sqrt2}(v+c_\xi H + s_\xi H'+iG_{Z_1})
\end{pmatrix} , \hs \chi = \frac{1}{\sqrt2}(\La -s_\xi H + c_\xi H'+iG_{Z_2}), \ee 
where $G_W\equiv \phi_1$, $G_{Z_1}\equiv A_1$, and $G_{Z_2}\equiv A_2$ are the massless Goldstone bosons eaten by $W$, $Z_1$, and $Z_2$ gauge bosons (see below), respectively. The field $H\equiv c_\xi S_1-s_\xi S_2$ is identical to the SM Higgs boson, while $H'\equiv s_\xi S_1+c_\xi S_2$ is a new Higgs boson associate to $U(1)_X\otimes U(1)_N$ symmetry breaking down to $U(1)_Y$. Their masses are given by
\bea m^2_H &=& \la_1 v^2 + \la_2\La^2-\sqrt{(\la_1 v^2 - \la_2\La^2)^2+\la_3^2v^2\La^2}\simeq \left(2\la_1-\frac{\la_3^2}{2\la_2}\right)v^2,\\
m^2_{H'} &=& \la_1 v^2 + \la_2\La^2+\sqrt{(\la_1 v^2 - \la_2\La^2)^2+\la_3^2v^2\La^2}\simeq 2\la_2\La^2,  \eea 
which are obviously that $m_H$ is at the weak scale, while $m_{H'}$ is at the $\La$ scale. The Higgs and new Higgs, i.e. $S_1$-$S_2$, mixing angle, $\xi$, is defined by
\be \tan (2\xi)\equiv t_{2\xi} = \frac{\la_3 v\La}{\la_2\La^2-\la_1v^2}\simeq \frac{\la_3}{\la_2}\frac{v}{\La},  \ee
which is small due to $v\ll \La$.

\subsection{Gauge sector}

Concerning gauge bosons, they acquire masses via the scalar kinetic term, $\sum_S(D^\mu S)^\dag (D_\mu S)$, when the symmetry breaking occurs. For charged gauge bosons, we obtain  
\be W^\pm = \frac{1}{\sqrt2}(A_1\mp iA_2), \ee
to be a physical field by itself, with mass
\be m^2_W = \fr 1 4 g^2v^2, \ee
 which is identical to that of the SM, given that $v=246$ GeV.

For neutral gauge bosons $(A_3,B,C)$, the fields $B$ and $C$ are generically not orthogonal and normalized due to the kinetic mixing term. Let us change to the canonical basis $(A_3,\hat{B},\hat{C})$ via a nonunitary transformation, $(A_3,\hat{B},\hat{C})^T=U_\delta (A_3,B,C)^T$, where
\be U_\delta = \begin{pmatrix}
1 & 0 & 0\\
0 & 1 & \delta\\
0 & 0 & \sqrt{1-\delta^2}
\end{pmatrix}. \ee

The mass matrix of neutral gauge bosons in the canonical basis, $(A_3,\hat{B},\hat{C})$, is given by
\be M_0^2 =\frac{g^2}{4} \begin{pmatrix}
v^2 & 0 & -\frac{t_Nv^2}{\sqrt{1-\delta^2}}\\
0 & 16t_X^2x^2\La^2 & -\frac{16t_Xt_Nx^2(1+\delta t)\La^2}{\sqrt{1-\delta^2}}\\
-\frac{t_Nv^2}{\sqrt{1-\delta^2}} & -\frac{16t_Xt_Nx^2(1+\delta t)\La^2}{\sqrt{1-\delta^2}} & \frac{t_N^2[v^2+16x^2(1+\delta t)^2\La^2]}{1-\delta^2}
\end{pmatrix}, \ee
where $t_X=g_X/g$, $t_N=g_N/g$, and $t=g_X/g_N$. This mass matrix has a zero eigenvalue (i.e., the photon mass) with a corresponding eigenstate (i.e., the photon field) to be
\be A = s_W A_3 + c_W (s_\theta \hat{B} + c_\theta \hat{C}), \ee
where the Weinberg's angle and the mixing angle $\theta$ are defined via $\tan$ functions, 
\be \tan (\theta_W)\equiv t_W = \frac{t_X}{\sqrt{1+2\delta t+t^2}},\hs \tan (\theta)\equiv t_\theta = \frac{1+\delta t}{t\sqrt{1-\delta^2}}. \ee
We define the SM $Z$ boson and the new gauge boson $Z'$ orthogonal to $A$, such as
\bea Z &=& c_W A_3 - s_W (s_\theta \hat{B} + c_\theta \hat{C}),\\
 Z' &=& c_\theta \hat{B} - s_\theta \hat{C}.\eea
 
 In the new basis $(A,Z,Z')$, the mass matrix $M^2_0$ is changed to 
 \be M^2 = U^T_W M_0^2 U_W = \begin{pmatrix}
0 & 0 & 0\\
0 & m^2_Z & m^2_{ZZ'}\\
0 & m^2_{ZZ'} & m^2_{Z'}
\end{pmatrix},\ee where
\be U_W = \begin{pmatrix}
s_W & c_W & 0\\
c_Ws_\theta & -s_Ws_\theta & c_\theta\\
c_Wc_\theta & -s_Wc_\theta & -s_\theta
\end{pmatrix}, \ee
and 
\bea m^2_Z &=& \frac{g^2}{4c^2_W} v^2,\label{Zmass}\\
m^2_{ZZ'} &=& \frac{g^2}{4c_W}t_Wt_\theta v^2,\\
m^2_{Z'} &=& \frac{g^2}{4}t^2_W\left[t^2_\theta v^2 +\frac{16t^2x^2(1-\delta^2)\La^2}{c^4_\theta}\right].
\eea
At this stage, the photon field $A$ is decoupled as a physical field, while the two states $Z$ and $Z'$ still mix by themselves via the $2\times 2$ mass matrix, as given. 

Finally, diagonalizing this $2\times 2$ mass matrix, we obtain two physical states,
\be Z_1 = c_\vph Z - s_\vph Z', \hs Z_2 = s_\vph Z + c_\vph Z', \ee 
with corresponding masses,
\bea
m^2_{Z_1} &=& \fr 1 2 \left[m^2_{Z}+m^2_{Z'}-\sqrt{(m^2_Z-m^2_{Z'})^2+4m^4_{ZZ'}}\right]\simeq m^2_Z-\fr{m^4_{ZZ'}}{m^2_{Z'}},\\
m^2_{Z_2} &=& \fr 1 2 \left[m^2_{Z}+m^2_{Z'}+\sqrt{(m^2_Z-m^2_{Z'})^2+4m^4_{ZZ'}}\right] \simeq m^2_{Z'}.
\eea
The $Z$-$Z'$ mixing angle is given by
\be t_{2\vph} = \frac{s_Ws_{2\theta}v^2}{(s_W^2s^2_\theta-c^2_\theta)v^2+16c_W^2t_X^2x^2\La^2}\simeq \frac{t\sqrt{1-\delta^2}(1+\delta t)}{8s_Wx^2(1+2\delta t+t^2)^2}\fr{v^2}{\La^2}. \ee The above approximations apply due to the fact that $v\ll \La$. Notice that the mixing effect between $Z$ and $Z'$ comes from two sources, the gauge symmetry breaking by $v,\La$ and the kinetic mixing term by $\delta$. Particularly, this mixing effect completely disappears when $\delta=-1/t = -g_N/g_X$; or in other words, two contributions to the mixing are canceled out.

Let us stress that the field $Z_1$ has a mass approximating that of the SM, called the SM $Z$-like boson, whereas the field $Z_2$ is a new gauge boson with a large mass at $\La$ scale. In addition, the $Z$-$Z'$ mixing angle is small due to $v\ll \La$. The observable effects in experiment concern only a shift in the $Z$ mass and the $Z$ couplings with SM particles, which are all suppressed by $(v/\La)^2$. Because of this suppression, the electroweak precision measurements on the $Z$ mass and the $Z$ couplings may be satisfied for a finite $\delta$ and $t=g_X/g_N$. 

Last, but not least, let us summarize the relation between the original gauge states and the physical eigenstates as $(A_3, B, C)^T = U_\delta^{-1} U_W U_\vph (A, Z_1, Z_2)^T$, where
\be U_\vph = \begin{pmatrix}
1 & 0 & 0\\
0 & c_\vph & s_\vph\\
0 & -s_\vph & c_\vph
\end{pmatrix}. \ee   

\section{\label{fgint} Fermion--gauge boson interaction}

Substituting the gauge bosons given in terms of physical fields above into the covariant derivative, we get
\bea D_\mu &=& \partial_\mu + ig_st_pG_{p\mu} + igs_WQA_\mu + ig(T_-W^-_\mu+\mathrm{H.c.})\crn
&&+\frac{ig}{c_W}\left\{c_\vph(T_3-s_W^2Q)-s_\vph s_W\left[X\fr{1+\delta t}{s_\theta c_\theta}-t_\theta Y\right]\right\} Z_{1\mu}\crn
&&+\frac{ig}{c_W}\left\{s_\vph(T_3-s_W^2Q)+c_\vph s_W\left[X\fr{1+\delta t}{s_\theta c_\theta}-t_\theta Y\right]\right\} Z_{2\mu}, \eea
where $T_{\pm}=(T_1\pm i T_2)/\sqrt2$ is weight-raising/lowering operator, respectively. The interactions of gauge bosons with fermions arise from the fermion kinetic term, i.e. $\sum_F\bar{F}i\gamma^\mu D_\mu F$.
It is easy to see that in the present model, gluons and photon interact with fermions similarly to the SM. In addition, the interaction of $W$ boson with quarks remains the same with the SM where the CKM matrix $V=V^\dag_{uL}V_{dL}$ enters the relevant quark vertex, whereas the interaction of $W$ boson with leptons is now modified by the PMNS matrix,
 \be \mathcal{L}^{\mathrm{CC}}_{\mathrm{int}}\supset -\frac{g}{\sqrt2}\bar{e}_{iL}\gamma^\mu U_{ij}\nu_{jL}W^-_\mu +\mathrm{H.c.},\ee
where the charged leptons are flavor diagonal, possessing a physical eigenstate index, $i$. 
 
Concerning the neutral currents coupled to $Z_{1,2}$, let us emphasize that $Q$, $T_3$, and $Y$ are universal for every flavors of neutrinos, charged leptons, up-type quarks, and down-type quarks. In addition, lepton flavors are universal under $X$, but quark flavors are not. Hence, the model predicts tree-level FCNCs for quarks, in addition to the flavor-conserving couplings. To extract the violating couplings, we look at quark-$Z_{1,2}$ interactions induced by $X$-charge, 
 \bea \mathcal{L}^{\mathrm{NC}}_{\mathrm{int}} &\supset& \frac{gt_W(1+\delta t)}{s_\theta c_\theta}\bar{q}_{L,R}\gamma^\mu V^\dag_{qL,R}\text{diag}(-x,-x,x)V_{qL,R}q_{L,R}(s_\vph Z_{1\mu}-c_\vph Z_{2\mu})\crn
&\supset &\frac{2xgt_W(1+\delta t)}{s_\theta c_\theta}\sum_{i\neq j}[V^*_{qL}]_{3i}[V_{qL}]_{3j}\bar{q}_{iL}\gamma^\mu q_{jL}(s_\vph Z_{1\mu}-c_\vph Z_{2\mu})+(L\to R), \eea which gives rise to the couplings of distinct $i$ and $j$ quarks, where $q$ denotes physical quarks of either up-type or down-type. As shown in \cite{VanDong:2022cin}, the new gauge boson $Z_2$ governs the FCNCs, while the $Z_1$ contribution is small and neglected.  
 
Furthermore, using unitarity conditions for lepton and quark mixing matrices $V_{lL,R}$ and $V_{qL,R}$, we obtain flavor-conserving interactions coupled to $Z_{1,2}$, given in the form of
\be \mathcal{L}^{\mathrm{NC}}_{\mathrm{int}}\supset-\frac{g}{2c_W}\{C_{\nu_L}^{Z_I}\bar{\nu}_{iL}\gamma^\mu\nu_{iL}+C_{\nu_R}^{Z_I}\bar{\nu}_{iR}\gamma^\mu\nu_{iR}+\bar{f}\gamma^\mu[g_V^{Z_I}(f)-g_A^{Z_I}(f)\gamma_5]f\}Z_{I\mu}, \ee
where $I=1,2$, while $f$ represents the physical charged fermions in the model, and
\bea C_{\nu_L}^{Z_1} &=& c_\vph-s_\vph s_W\left[\fr{2x(1+\delta t)}{s_\theta c_\theta}+t_\theta \right], \\
C_{\nu_R}^{Z_1} &=& -\frac{2s_\vph s_W x(1+\delta t)}{s_\theta c_\theta} , \\ C_{\nu_{L,R}}^{Z_2} &=& C_{\nu_{L,R}}^{Z_1}|_{c_\vph\to s_\vph,s_\vph\to -c_\vph}, \eea
and \bea 
g_V^{Z_1}(f) &=& c_\vph[T_3(f_L)-2s_W^2Q(f)]\crn
&&-s_\vph s_W\left\{t_\theta[T_3(f_L)-2Q(f)]+2X(f)\frac{1+\delta t}{s_\theta c_\theta}\right\}, \\
g_A^{Z_1}(f) &=& T_3(f_L)(c_\vph-s_\vph s_W t_\theta), \\
g_{V,A}^{Z_2}(f) &=& g_{V,A}^{Z_1}(f)|_{c_\vph\to s_\vph, s_\vph\to -c_\vph}.
\eea For later usage, we collect the couplings of $Z_1$ and $Z_2$ with charged fermions in Tables \ref{tab2} and \ref{tab3}, respectively. It is clear that in the limit $\vph\to 0$, the $Z_1$ couplings with ordinary fermions are identical to those of the SM.

\begin{table}[h]
\bc
\begin{tabular}{l|cc}
\hline\hline
$f$ & $g^{Z_1}_V(f)$ & $g^{Z_1}_A(f)$ \\
\hline 
$e,\mu,\tau$ & $c_\vph\left(2s^2_W-\fr 1 2\right)-s_\vph s_W \left[\fr{3}{2}t_\theta+\fr{2x(1+\delta t)}{s_\theta c_\theta}\right]$ & $\fr 1 2\left(s_\varphi s_W t_\theta - c_\varphi\right)$\\
$u,c$ & $c_\vph\left(\fr 1 2-\fr 4 3 s^2_W\right)+ s_\vph s_W\left[\fr{5}{6}t_\theta+\fr{2x(1+\delta t)}{s_\theta c_\theta}\right]$ & $\fr 1 2\left(c_\varphi-s_\varphi s_W t_\theta\right)$\\
$t$ & $c_\vph\left(\fr 1 2-\fr 4 3 s^2_W\right)+ s_\vph s_W\left[\fr{5}{6}t_\theta-\fr{2x(1+\delta t)}{s_\theta c_\theta}\right]$ & $\fr 1 2\left(c_\varphi-s_\varphi s_W t_\theta\right)$\\
$d,s$ & $c_\vph\left(\fr 2 3 s^2_W-\fr 1 2\right)- s_\vph s_W\left[\fr{1}{6}t_\theta-\fr{2x(1+\delta t)}{s_\theta c_\theta}\right]$ & $\fr 1 2\left(s_\varphi s_W t_\theta - c_\varphi\right)$\\
$b$ & $c_\vph\left(\fr 2 3 s^2_W-\fr 1 2\right)- s_\vph s_W\left[\fr{1}{6}t_\theta+\fr{2x(1+\delta t)}{s_\theta c_\theta}\right]$ & $\fr 1 2\left(s_\varphi s_W t_\theta - c_\varphi\right)$\\
\hline\hline
\end{tabular}
\caption[]{\label{tab2} Couplings of $Z_1$ with charged fermions.}
\ec
\end{table}

\begin{table}[h]
\bc
\begin{tabular}{l|cc}
\hline\hline
$f$ & $g^{Z_2}_V(f)$ & $g^{Z_2}_A(f)$ \\
\hline 
$e,\mu,\tau$ & $s_\vph\left(2s^2_W-\fr 1 2\right)+c_\vph s_W \left[\fr{3}{2}t_\theta+\fr{2x(1+\delta t)}{s_\theta c_\theta}\right]$ & $-\fr 1 2\left(c_\varphi s_W t_\theta + s_\varphi\right)$\\
$u,c$ & $s_\vph\left(\fr 1 2-\fr 4 3 s^2_W\right)-c_\vph s_W\left[\fr{5}{6}t_\theta+\fr{2x(1+\delta t)}{s_\theta c_\theta}\right]$ & $\fr 1 2\left(s_\varphi+c_\varphi s_W t_\theta\right)$\\
$t$ & $s_\vph\left(\fr 1 2-\fr 4 3 s^2_W\right)-c_\vph s_W\left[\fr{5}{6}t_\theta-\fr{2x(1+\delta t)}{s_\theta c_\theta}\right]$ & $\fr 1 2\left(s_\varphi+c_\varphi s_W t_\theta\right)$\\
$d,s$ & $s_\vph\left(\fr 2 3 s^2_W-\fr 1 2\right)+c_\vph s_W\left[\fr{1}{6}t_\theta-\fr{2x(1+\delta t)}{s_\theta c_\theta}\right]$ & $-\fr 1 2\left(c_\varphi s_W t_\theta + s_\varphi\right)$\\
$b$ & $s_\vph\left(\fr 2 3 s^2_W-\fr 1 2\right)+c_\vph s_W\left[\fr{1}{6}t_\theta+\fr{2x(1+\delta t)}{s_\theta c_\theta}\right]$ & $-\fr 1 2\left(c_\varphi s_W t_\theta + s_\varphi\right)$\\
\hline\hline
\end{tabular}
\caption[]{\label{tab3} Couplings of $Z_2$ with charged fermions.}
\ec
\end{table}

\section{\label{constraint} Experimental constraints}

In the model under consideration, the SM $Z$ boson mixes with the new neutral gauge boson $Z'$ through the kinetic mixing and the symmetry
breaking, which gives rise to interesting phenomena. Notice that in the original framework with two Higgs doublets, some bounds were given, but suppressing the kinetic mixing effect \cite{VanDong:2022cin}. In this section, we will deliver a full analysis to various new physics processes and relevant constraints.
 
\subsection{\label{Wmass1}$W$-boson mass deviation}
Recently, the CDF collaboration has reported a new result of $W$-boson mass, $m_W|_{\mathrm{CDF}}=80.5335\pm 0.0094$ GeV \cite{CDF:2022hxs}, which is $7\sigma$ above the SM prediction of $m_W|_\mathrm{SM}=80.357\pm 0.006$ GeV \cite{ParticleDataGroup:2020ssz}. This may be a significant indication of new physics beyond the SM. 

In present model, because of the $Z$-$Z'$ mixing, the observed $Z_1$-boson mass is reduced in comparison with the SM $Z$-boson mass, while the mass of $W$ boson is retained. This gives rise to a positive contribution to the Peskin-Takeuchi $T$-parameter at tree-level,
\be \al T=\rho -1= \frac{m^2_W}{c^2_Wm^2_{Z_1}}-1 = \frac{m^2_Z}{m^2_{Z_1}}-1\simeq \frac{m^4_{ZZ'}}{m^2_Zm^2_{Z'}}\simeq \frac{(1+\delta t)^2}{16x^2(1+2\delta t+t^2)^2}\frac{v^2}{\La^2}.\ee
Since the mass of $Z_1$ boson is already fixed, the positive value of $\al T$ dominantly enhances the mass of $W$ boson (cf. \cite{Peskin:1991sw,Strumia:2022qkt}), such as
\be \Delta m^2_W = \frac{c^4_W m^2_Z}{c^2_W-s^2_W}\al T\simeq \frac{g^2c_W^2(1+\delta t)^2}{64 c_{2W}x^2(1+2\delta t+t^2)^2}\frac{v^4}{\La^2}.\ee

Taking the central values of $W$ mass from the CDF experiment and the SM prediction, as well as using $v=246$ GeV, $s^2_W=0.231$ and $\al=1/128$, we obtain
\be \La \simeq 1.68\times \left|\frac{1+\delta t}{x(1+2\delta t+t^2)}\right|\text{ TeV}. \label{Wmass}\ee 

\subsection{\label{decaywidth}Total $Z_1$ decay width}

The $Z$-$Z'$ mixing leads to a deviation in couplings of $Z_1$ with SM fermions, thus causing a shift of the total $Z_1$ decay width, compared with the SM prediction. For convenience, we rewrite the Lagrangian describing the $Z_1$ couplings with SM fermions as follows
\be \mathcal{L}_{\mathrm{int}}\supset-\frac{g}{2c_W}\{\bar{\nu}_{iL}\gamma^\mu(1+\Delta_{\nu_L})\nu_{iL}+\bar{f}\gamma^\mu[g_V^Z(f)(1+\Delta^V_f)-g_A^Z(f)(1+\Delta^A_f)\gamma_5]f\}Z_{1\mu}, \label{LaZin}\ee
where $g_V^Z(f)=T_3(f)-2Q(f)s_W^2$ and $g_A^Z(f)=T_3(f)$ are the vector and axial-vector couplings predicted by the SM, while $\Delta_{\nu_L}$, $\Delta^V_f$, and $\Delta^A_f$ are the coupling deviations predicted by the present model, evaluated by 
\bea
\Delta_{\nu_L} &\simeq& -\frac{(1+\delta t)[1+\delta t+2x(1+2\delta t+t^2)]}{16x^2(1+2\delta+t)^2t^2}\frac{v^2}{\La^2},\\
\Delta^V_f &\simeq& -\frac{(1+\delta t)\{[T_3(f)-2Q(f)](1+\delta t)+2X(f)(1+2\delta t+t^2)\}}{16[T_3(f)-2s^2_WQ(f)]x^2(1+2\delta+t)^2t^2}\frac{v^2}{\La^2},\\
\Delta^A_f &\simeq& -\frac{(1+\delta t)^2}{16x^2(1+2\delta+t)^2t^2}\frac{v^2}{\La^2}.
\eea

Using the Lagrangian (\ref{LaZin}), we derive the total decay width of $Z_1$ in the form as $\Gamma_{Z_1} = \Gamma_Z+\Delta\Gamma_Z$, where $\Gamma_Z$ is the SM part, while the shift $\Delta\Gamma_Z$ is given by
\bea \Delta\Gamma_Z &\simeq& \frac{m_Z}{6\pi}\left(\frac{g}{2c_W}\right)^2\left\{\sum_f N_C(f)\left[\left(g_V^Z(f)\right)^2\Delta^V_f+\left(g_A^Z(f)\right)^2\Delta^A_f\right]+\frac{3\Delta_{\nu_L}}{2}\right\}\crn
&& +\frac{\Delta m_Z}{12\pi}\left(\frac{g}{2c_W}\right)^2\left\{\sum_f N_C(f)\left[\left(g_V^Z(f)\right)^2+\left(g_A^Z(f)\right)^2\right]+\frac{3}{2}\right\}, \eea
where $N_C(f)$ is the color number of the fermion $f$, $m_Z$ is determined as in Eq. (\ref{Zmass}), and
\be \Delta m_Z\equiv m_{Z_1}-m_Z\simeq -\frac{g}{2c_W}\frac{(1+\delta t)^2}{32x^2(1+2\delta t+t^2)^2}\frac{v^3}{\La^2}  .\ee

Comparing the experimental measurement of the total $Z_1$ decay width, $\Gamma^{\mathrm{exp}}_{Z_1} = 2.4955\pm 0.0023$ GeV, with the SM prediction, $\Gamma^{\mathrm{SM}}_Z = 2.4941\pm 0.0009$~GeV, leads to $|\Delta\Gamma_Z|<0.0046$~GeV \cite{ParticleDataGroup:2022pth}, which subsequently constrains 
\be \La\gtrsim \sqrt{\left|\frac{1.56(1+\delta t)+x(1+2\delta t+t^2)}{0.61x^2t^2(1+2\delta+t)^2(1+\delta t)^{-1}}+\frac{(1+\delta t)^2}{x^2(1+2\delta t+t^2)^2}\right|} \text{ TeV}.\ee

\subsection{\label{mixing}$Z$-$Z'$ mixing angle}

The size of the $Z$-$Z'$ mixing angle, $\vph$, is strongly constrained by the precision $Z$-pole experiments at LEP and SLC \cite{ALEPH:2005ab}. This new physics effect is safe if one imposes the mixing parameter to be $|\varphi|\lesssim 10^{-3}$ \cite{Erler:2009jh}. This leads to
\be \La\gtrsim 2.81\frac{\sqrt{t|1+\delta t|\sqrt{1-\delta^2}}}{|x(1+2\delta t+t^2)|}\text{ TeV}. \ee

\subsection{\label{LEPII}LEPII}

Because the $Z$-$Z'$ mixing is strongly suppressed by $v^2/\La^2$, we can omit this mixing effect for the processes above the $Z$ pole, hereafter. Namely, taking $\varphi\to 0$, thus $Z_2\to Z'$, results in $Z'$ to be a physical field. 
With the mass at TeV scale, $Z'$ is certainly not produced in electron-positron collisions at LEPII experiment \cite{ALEPH:2006bhb}. However, it can indirectly modify observables, which deviate from the SM predictions.

One of the important processes at LEPII is the pair production of leptons $e^+e^-\to \bar{f}f$ with $f=e,\mu,\tau$ \cite{ALEPH:2013dgf}. Such processes can be induced by the $Z'$ exchange, which is described by the effective Lagrangian as follows
\be \mathcal{L}_{\text{eff}} = \frac{1}{1+\delta_{ef}}\fr{g^2}{c^2_Wm_{Z'}^2}\sum_{i,j=L,R}C_i^{Z'}(e)C_j^{Z'}(f) \bar{e}_i\gamma_\mu e_i\bar{f}_j\gamma^\mu f_j, \ee
where $\delta_{ef}=1(0)$ for $f=e(f\neq e)$ and $C_{L,R}^{Z'}(f)=\frac{1}{2}[g_V^{Z_2}(f)\pm g_A^{Z_2}(f)]$. Considering the case of dimuon in final state, we have
\be \mathcal{L}_{\text{eff}} \supset \fr{g^2[C_L^{Z'}(e)]^2}{c^2_Wm_{Z'}^2}\bar{e}_L\gamma_\mu e_L\bar{\mu}_L\gamma^\mu \mu_L+(LR)+(RL)+(RR), \ee
where the last three terms differ from the first one only in chiral structures. 

Taking a typical bound for the models like ours \cite{Carena:2004xs},
\be \fr{g^2[C_L^{Z'}(e)]^2}{c^2_Wm_{Z'}^2}<\fr{1}{(6\text{ TeV})^2}, \ee
we obtain
\be \Lambda\gtrsim 3\times\left|\frac{0.5(1+\delta t)+x(1+2\delta t+t^2)}{x(1+2\delta t+t^2)}\right|\text{ TeV}. \ee

\subsection{\label{LHC}LHC}

At the LHC, the new gauge boson $Z'$ can be resonantly produced via quark fusion $\bar{q}q\to Z'$ and subsequently decayed to two-jet (dijet) or two-lepton (dilepton) final states, in which the most significant decay channel is $Z'\to \bar{l}l$ with $l=e,\mu$. Using narrow width approximation, the cross-section for producing a dilepton final state can be estimated as
\be \sigma(pp\to Z'\to l \bar{l})\simeq \frac{\pi}{3}\left(\frac{g}{2c_W}\right)^2\sum_{q}L_{q\bar{q}}(m^2_{Z'})\left\{\left[g_V^{Z_2}(q)\right]^2+\left[g_A^{Z_2}(q)\right]^2\right\}\fr{\Ga(Z'\to \bar{l}l)}{\Ga_{Z'}},  \ee  
with the parton luminosities $L_{q\bar{q}}$ given by
\begin{eqnarray}
L_{q\bar{q}}(m^2_{Z'})=\int^1_{\frac{m^2_{Z'}}{s}}\frac{dx}{xs}\left[f_q(x,m^2_{Z'})f_{\bar{q}}\left(\frac{m^2_{Z'}}{xs},m^2_{Z'}\right)+f_q\left(\frac{m^2_{Z'}}{xs},m^2_{Z'}\right)f_{\bar{q}}(x,m^2_{Z'})\right],
\end{eqnarray}
where $\sqrt{s}$ is the collider center-of-mass energy, and $f_{q(\bar{q})}(x,m^2_{Z'})$ is the parton distribution function of quark $q$ (antiquark $\bar{q}$), evaluated at the scale $m_{Z'}$. Additionally, the partial decay width of $Z'$ is given by 
\be
 \Ga(Z'\to \bar{l}l)\simeq\fr{g^2m_{Z'}}{48\pi c^2_W}\left\{[g^{Z_2}_V(l)]^2+[g^{Z_2}_A(l)]^2\right\},\ee
 while the total $Z'$ decay width is estimated as
 \bea
 \Ga_{Z'} &\simeq&\fr{g^2m_{Z'}}{48\pi c^2_W}\sum_{f} N_C(f) \left\{[g^{Z_2}_V(f)]^2+[g^{Z_2}_A(f)]^2\right\}\crn
 &&+\frac{g^2m_{Z'}}{32\pi c^2_W}(C^{Z_2}_{\nu_L})^2 +\frac{m_{Z'}}{192\pi m^2_Z}(C_{Z'ZH})^2\crn
 &&+\sum_{i=1,2,3} \frac{g^2m_{Z'}}{96\pi c^2_W}(C^{Z_2}_{\nu_R})^2\left(1-\frac{4m^2_{\nu_{iR}}}{m^2_{Z'}}\right)^{3/2}\Theta\left(\frac{m_{Z'}}{2}-m_{\nu_{iR}}\right),\eea 
 given that $m_{Z'}<m_{H'}$. Here, $f$ denotes charged fermions, $N_C(f)$ is the color number of $f$, the $Z'ZH$ coupling is given by $C_{Z'ZH}\simeq \frac{g^2v}{2c_W}t_Wt_\theta$, and $\Theta$ is the step function.
 
\begin{figure}[!h]
\begin{center}
	\includegraphics[scale=0.4]{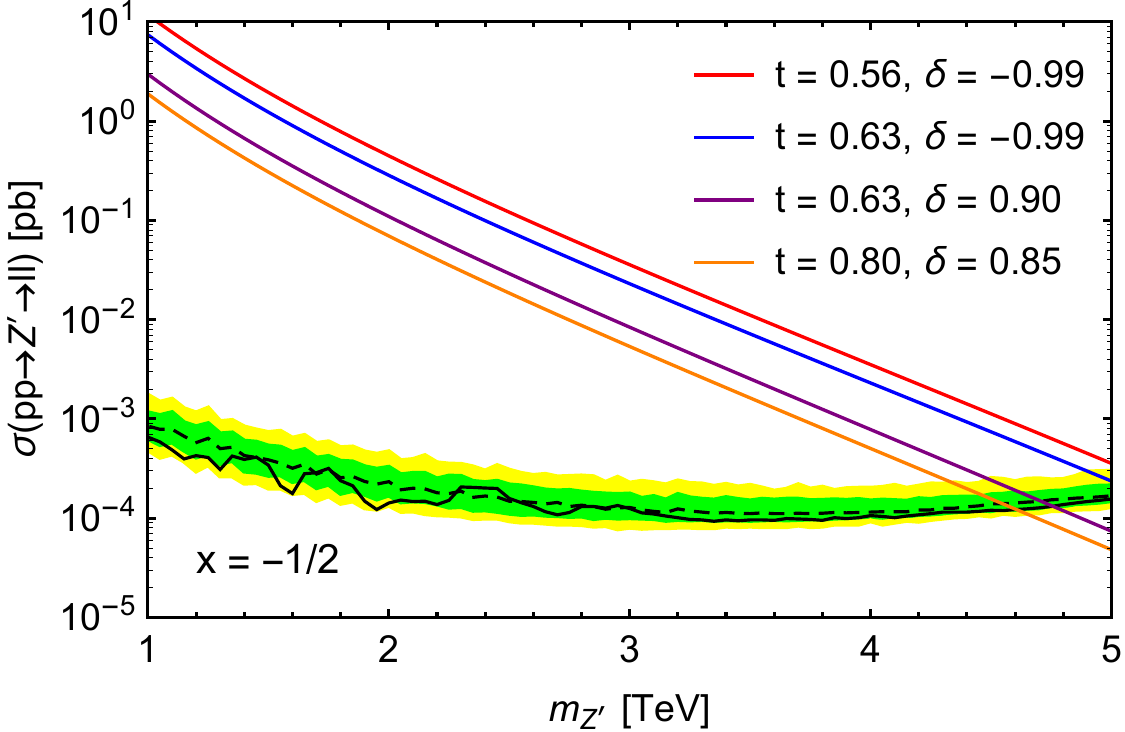}
	\includegraphics[scale=0.4]{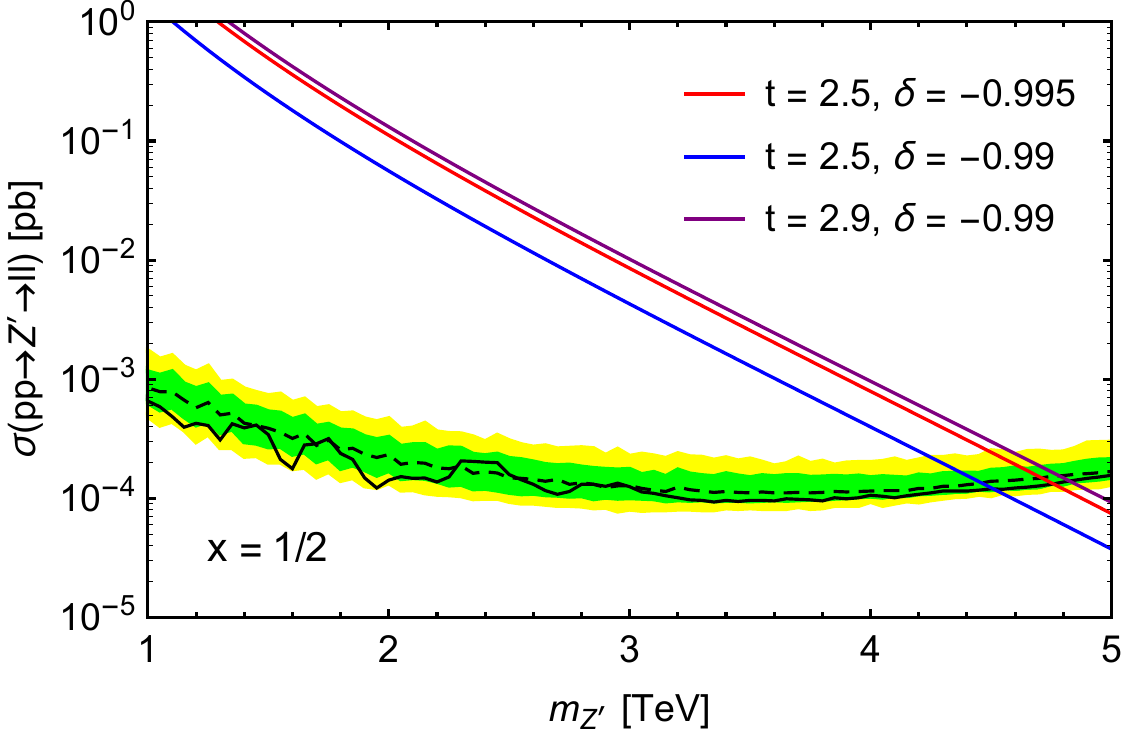}
	\includegraphics[scale=0.4]{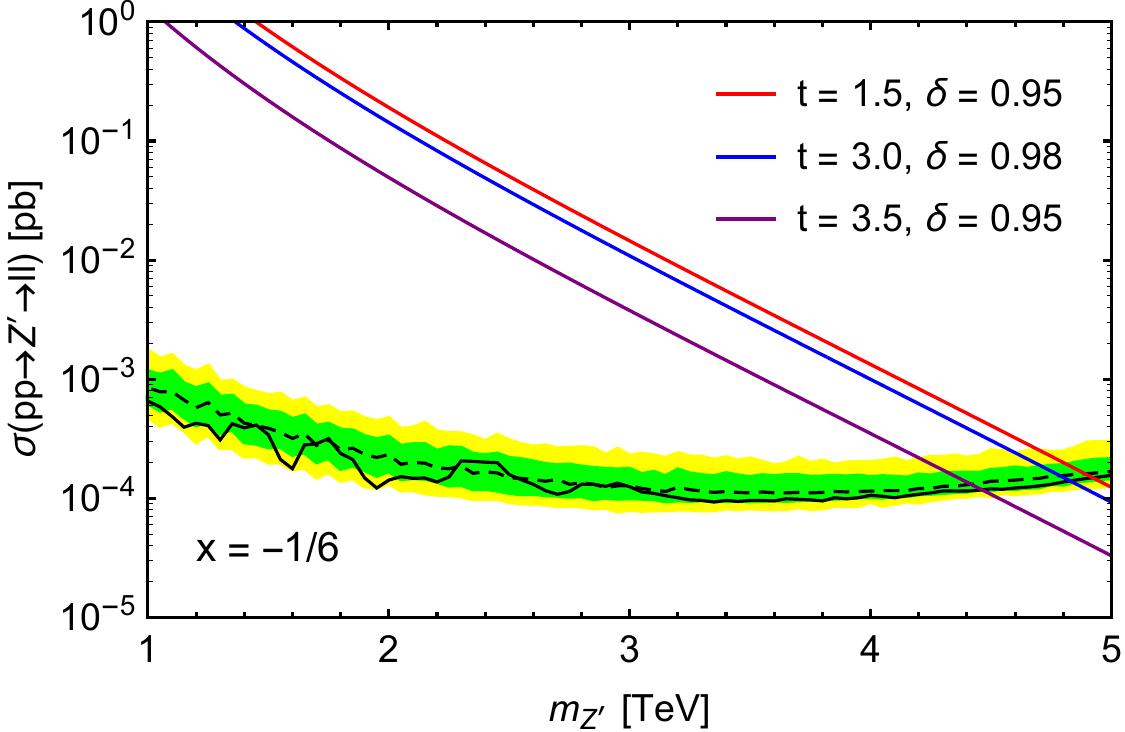}
	\includegraphics[scale=0.4]{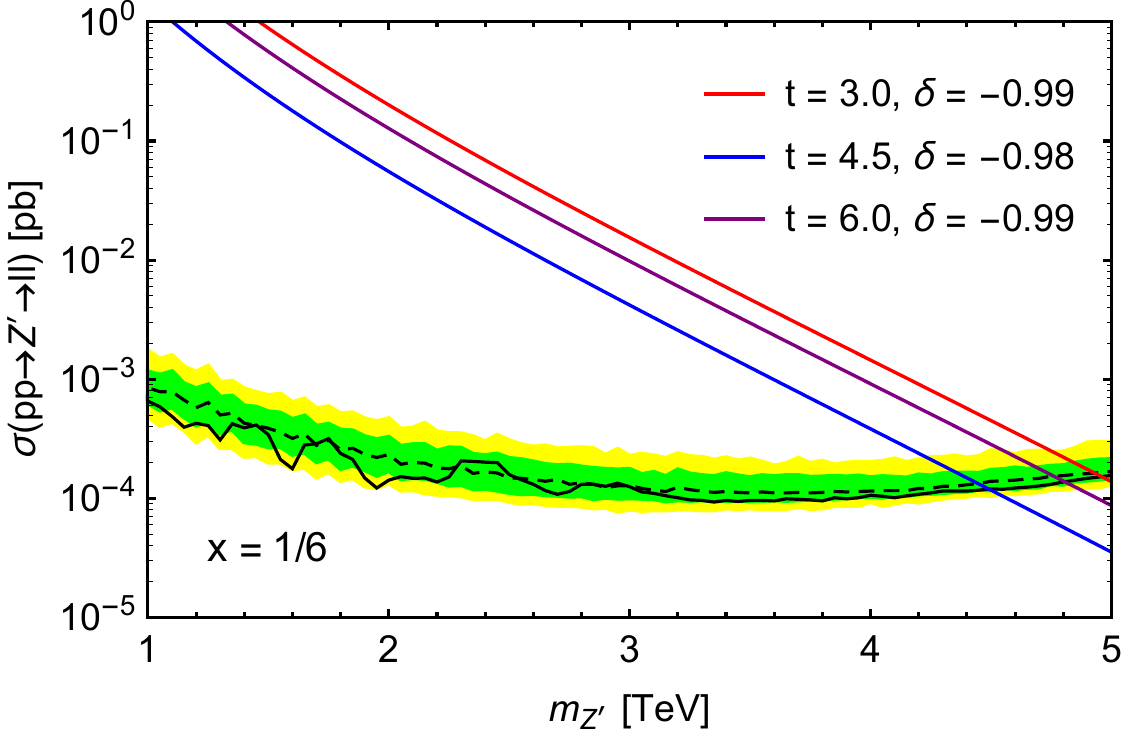}
\caption{\label{fig1}The cross section for process $\sigma(pp\to Z'\to \bar{l}l)$ as a function of $Z'$-boson mass according to various values of $t$ and $\delta$ for each fixed value of $x$. The solid and dashed black curves refer to the observed and expected limits, while the green and yellow bands refer to $1\sigma$ and $2\sigma$ expected limits, respectively \cite{Aaboud:2017buh}.}	
	\end{center}
\end{figure}

 In Fig. \ref{fig1}, we plot the dilepton production cross section $\sigma(pp\to Z'\to \bar{l}l)$ as a function of new gauge boson mass, $m_{Z'}$, according to various values of $t$ and $\delta$ for each fixed value of $x$, assuming $m_{\nu_{1R}}=m_{\nu_{2R}}= m_{Z'}/3$, and $m_{\nu_{3R}}\simeq m_{Z'}/2$. The selected values of $t$ and $\delta$ are within allowed region, satisfying all above experimental bounds (as also shown in Fig. \ref{fig2} below). Let us stress that in the model under consideration, the parameter $x$ can take an arbitrary non-zero value as far as the final unification has been not ascertained. In the current numerical analysis, the values of $x$ may be chosen as in \cite{VanDong:2022cin}, without loss of generality. In addition, we include upper limits for the cross section of this process at 95\% credibility level using 36.1 fb$^{-1}$ of $p\,p$ colision at $\sqrt{s}=13$ TeV by ATLAS experiment \cite{Aaboud:2017buh}. From this figure, it indicates that the lower bound for new gauge boson mass is generally around $4.5$ TeV. 

In Fig. \ref{fig2}, we combine the lower bounds on the new physics scale, which come from the LHC (brown curve), the LEPII (orange curve), the $Z$-$Z'$ mixing (blue curve), and the $Z_1$ decay width (purple curve). The available regions for the new physics scale lie above these four curves. Note that the incompleteness of the brown curves is due to a limitation of the current LHC data \cite{Aaboud:2017buh}. Additionally, for each case, we plot a red curve based on Eq. (\ref{Wmass}) concerning the $W$-mass deviation measured by the CDF collaboration. We thus achieve allowed parameter spaces as collected in Table \ref{tab4}, which are directly extracted from Fig. \ref{fig2}.

\begin{figure}[!h]
\begin{center}
	\includegraphics[scale=0.28]{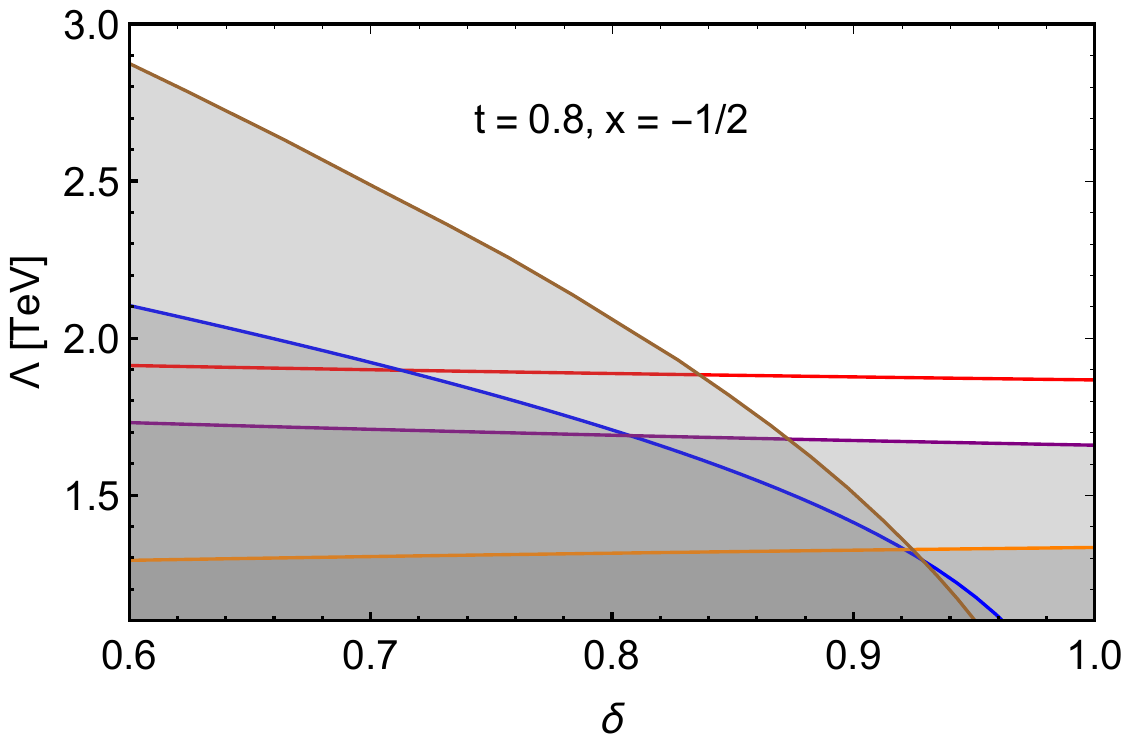}
	\includegraphics[scale=0.28]{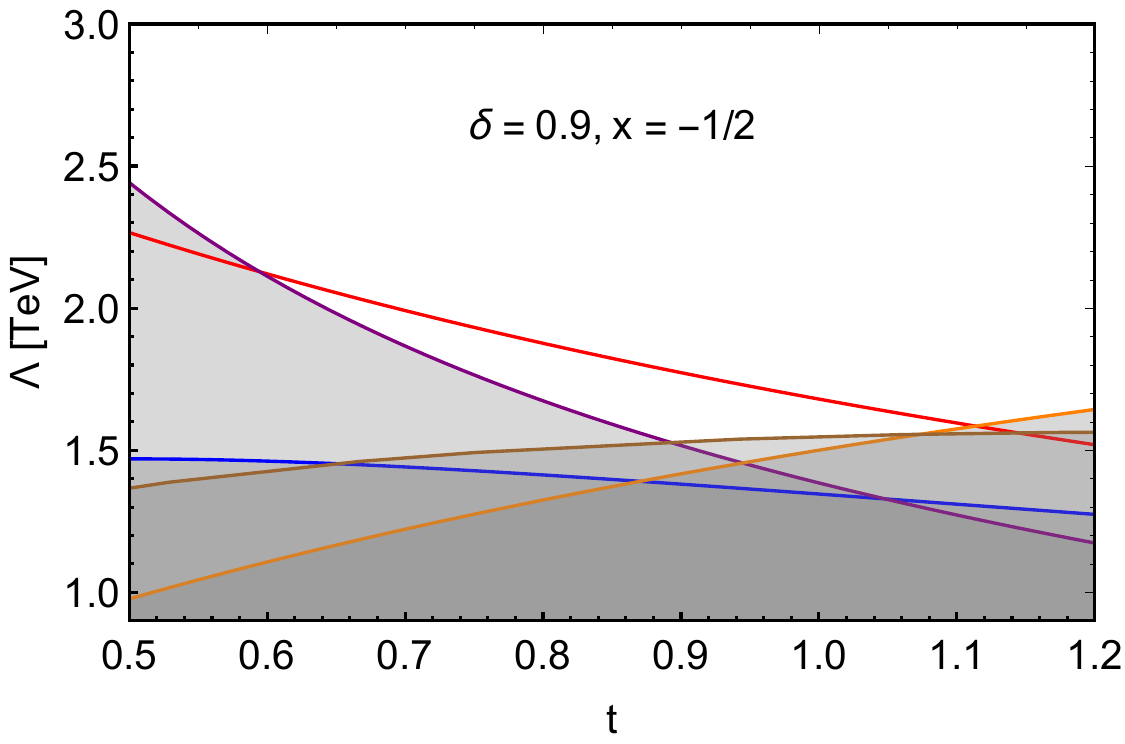}
	\includegraphics[scale=0.28]{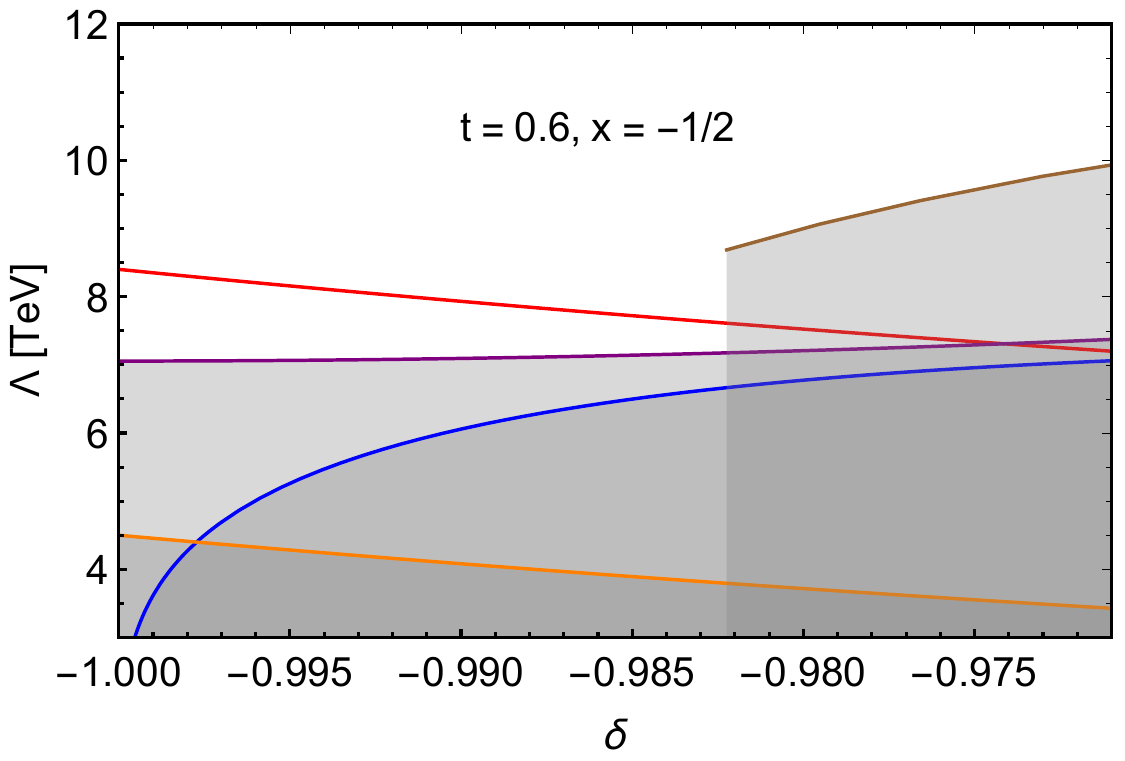}
	\includegraphics[scale=0.28]{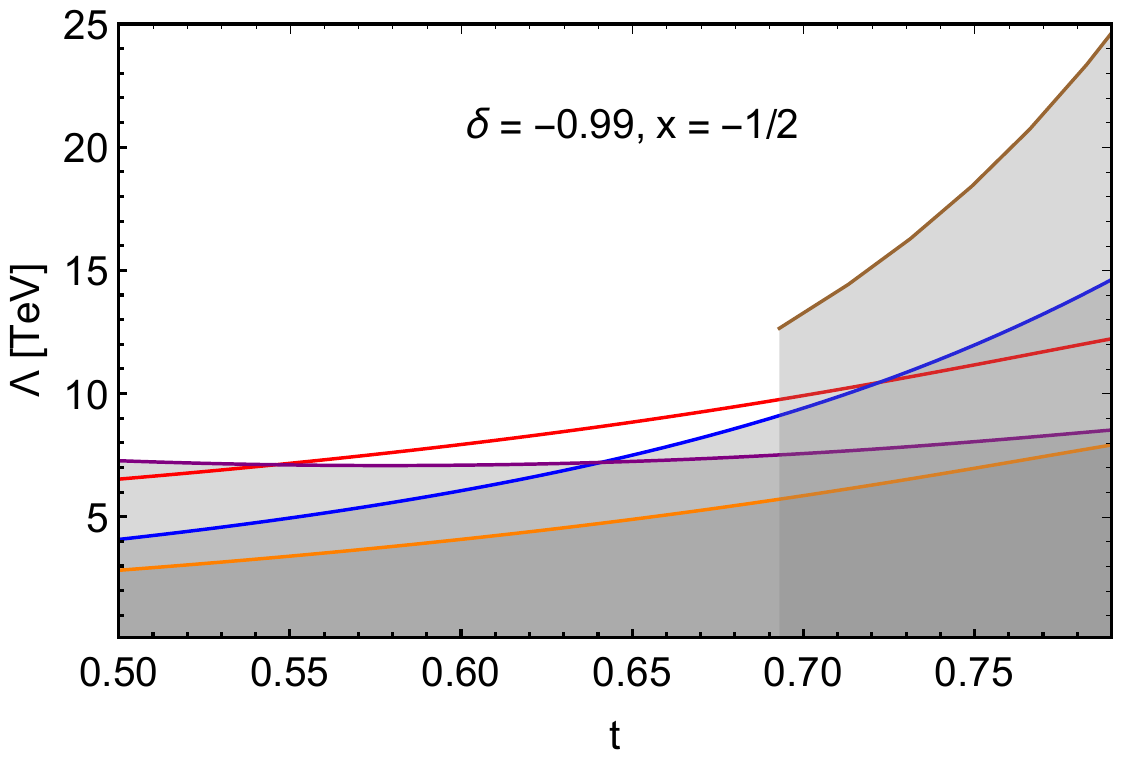}
	\includegraphics[scale=0.28]{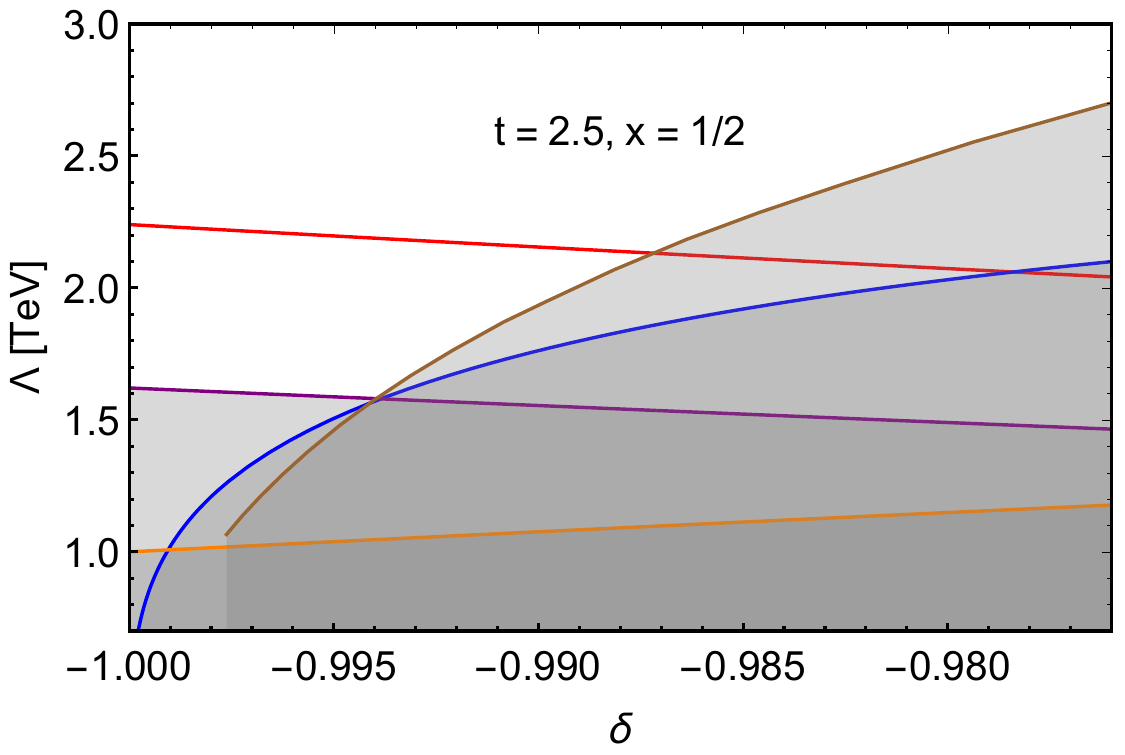}
	\includegraphics[scale=0.28]{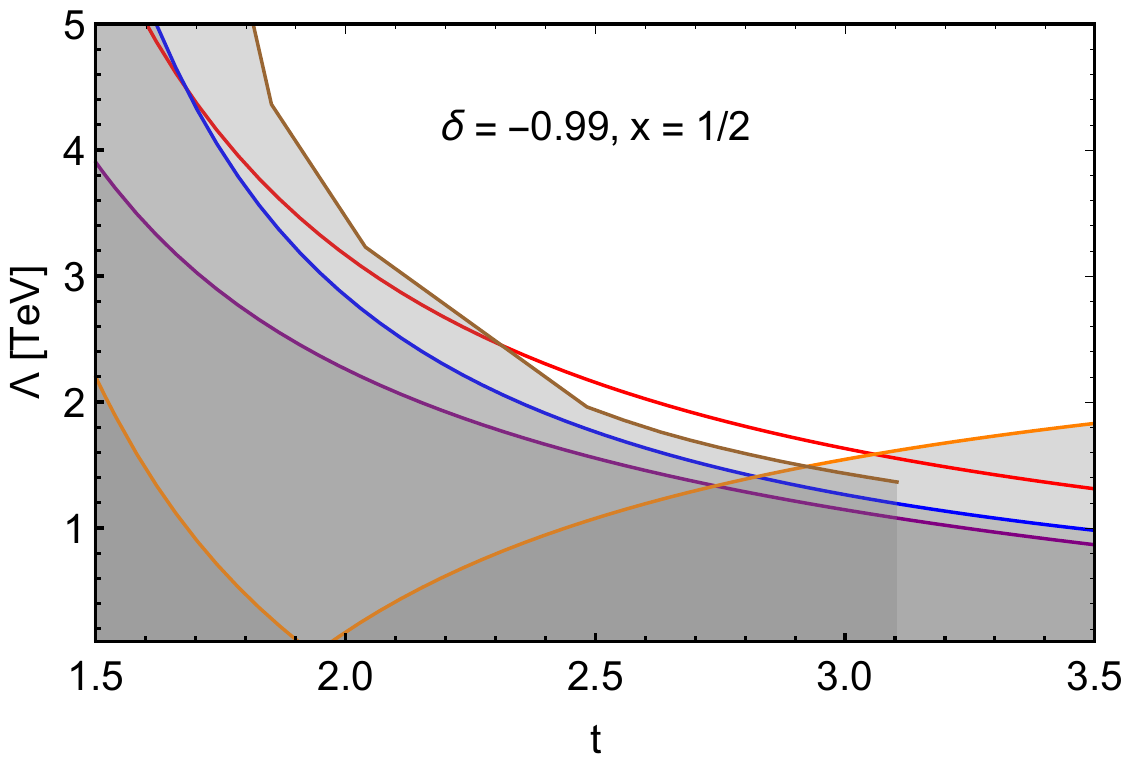}
	\includegraphics[scale=0.28]{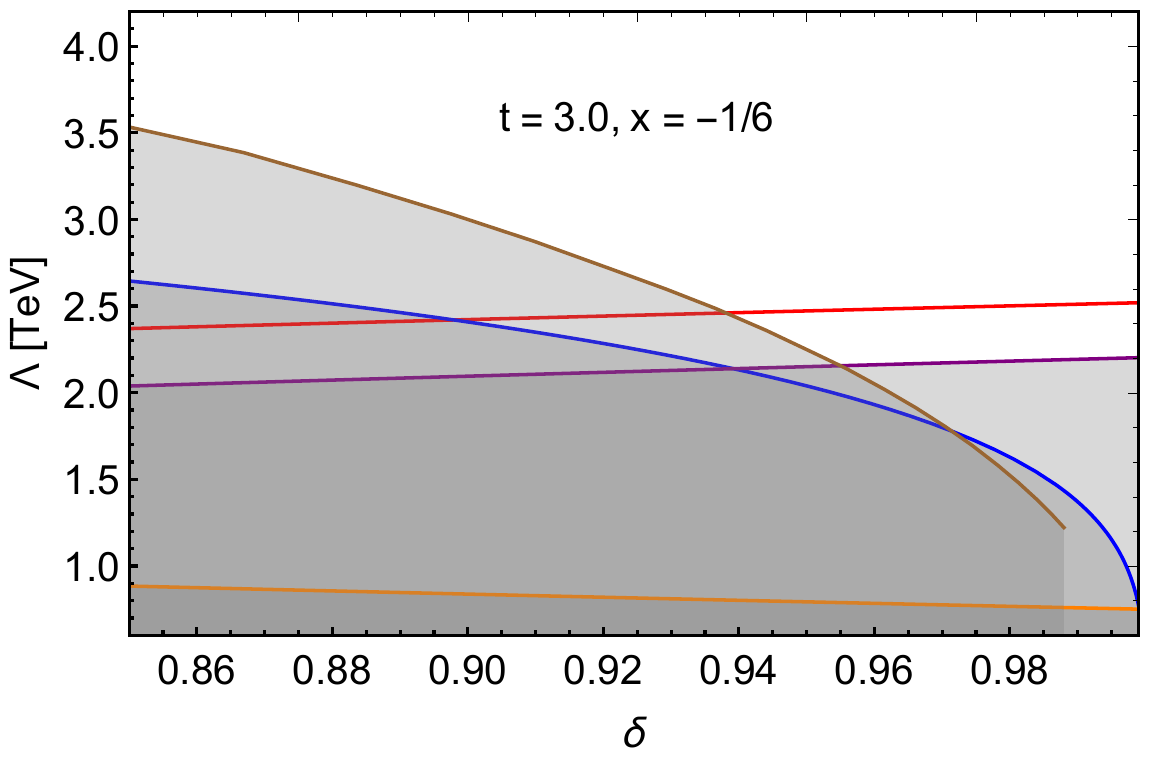}
	\includegraphics[scale=0.28]{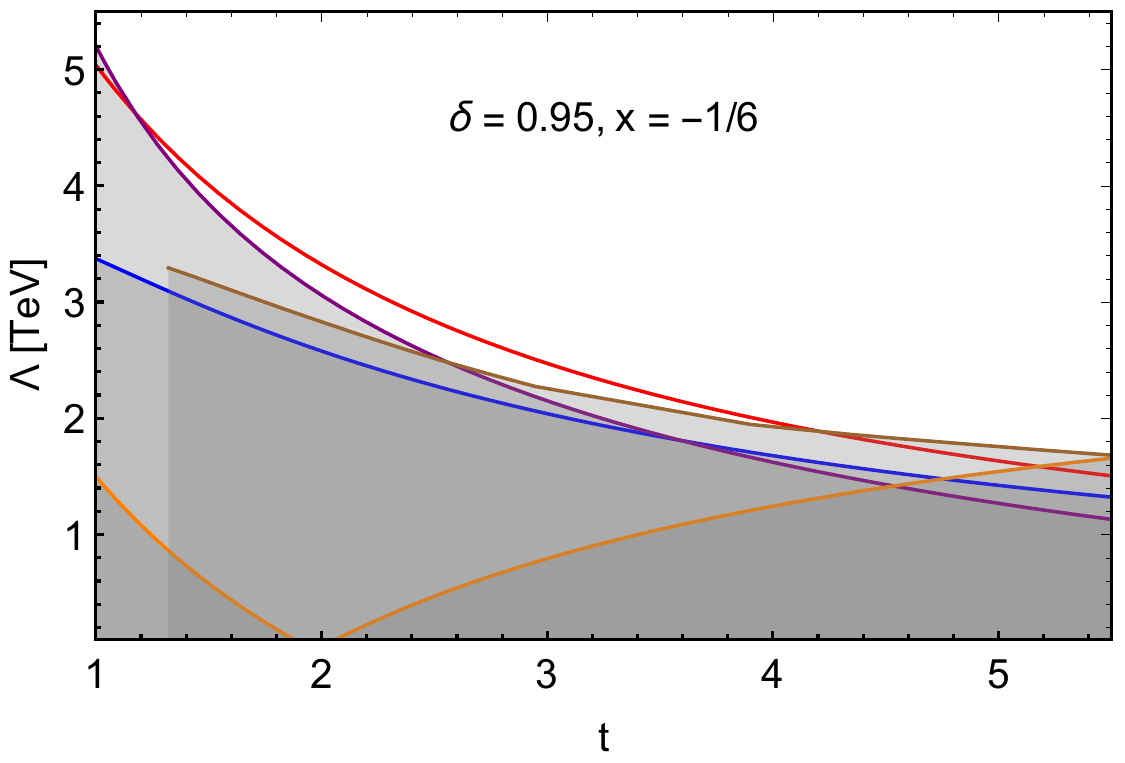}
	\includegraphics[scale=0.28]{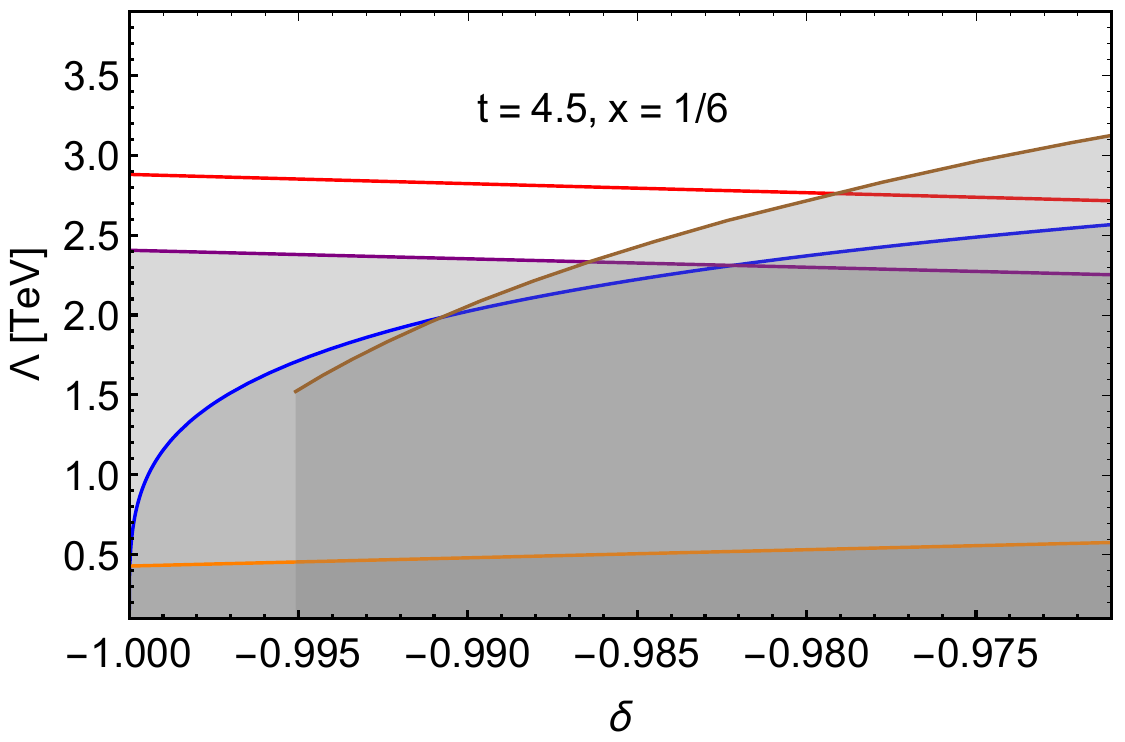}
	\includegraphics[scale=0.28]{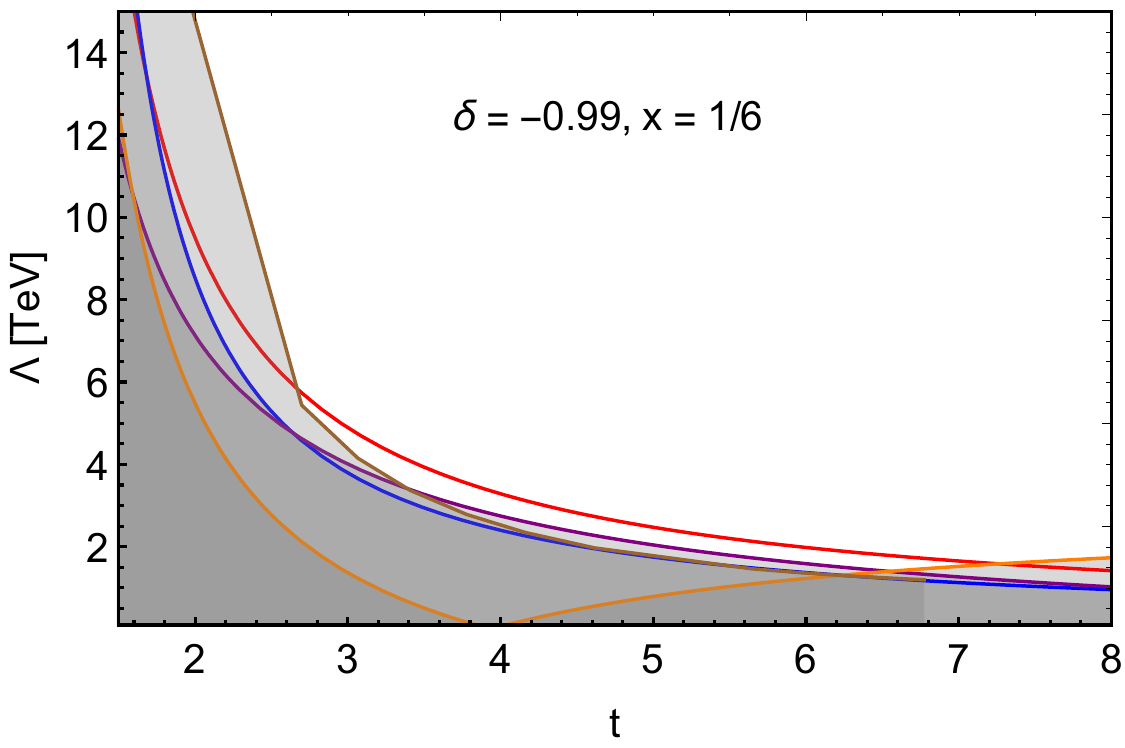}
\caption{\label{fig2}The brown, orange, blue, and purple curves denote the lower bounds on the new physics scale obtained from the current LHC limit for dilepton production, the LEPII dilepton signal constraint, the $Z$-$Z'$ mixing parameter, and the precision measurement of $Z_1$ decay width, respectively. The shaded regions are excluded. The red curves are based on Eq. (\ref{Wmass}), which results from the CDF precision measurement on the $W$-boson mass.}	
	\end{center}
\end{figure}

\begin{table}[h]
\bc
\begin{tabular}{ccc|ccc}
\hline\hline
$(x,t)$ & $\delta$ & $\Lambda$ [TeV] & $(x,\delta)$ & $t$ & $\Lambda$ [TeV] \\ \hline
$(-1/2,0.8)$ & $0.836\lesssim\delta< 1$ & $1.86\lesssim\Lambda\lesssim 1.88$ &  $(-1/2,0.90)$ & $0.59\lesssim t \lesssim 1.11$ & $1.59\lesssim\Lambda\lesssim 2.13$\\
$(-1/2,0.6)$ & $-1<\delta\lesssim -0.982$ & $7.60\lesssim\Lambda\lesssim 8.39$ &  $(-1/2,-0.99)$ & $0.55\lesssim t\lesssim 0.69$ & $7.13\lesssim\Lambda\lesssim 9.76$\\
$(1/2,2.5)$ & $-1<\delta\lesssim -0.987$ & $2.13\lesssim\Lambda\lesssim 2.24$ &  $(1/2,-0.99)$ & $2.31\lesssim t\lesssim 3.06$ & $1.58\lesssim\Lambda\lesssim 2.45$\\
$(-1/6,3.0)$ & $0.94\lesssim\delta<1$ & $2.45\lesssim\Lambda\lesssim 2.52$ &  $(-1/6,0.95)$ & $1.19\lesssim t\lesssim 4.20$ & $1.89\lesssim\Lambda\lesssim 4.61$\\
$(1/6,4.5)$ & $-1<\delta\lesssim -0.979$ & $2.76\lesssim\Lambda\lesssim 2.88$ &  $(1/6,-0.99)$ & $2.68\lesssim t\lesssim 7.27$ & $1.56\lesssim\Lambda\lesssim 5.83$\\
\hline\hline
\end{tabular}
\caption[]{\label{tab4} Parameter spaces of the model satisfy all current experiential constraints, including the LHC, the LEPII, electroweak precision measurements, and $W$-boson mass deviation.}  
\ec
\end{table}

\subsection{Non-DM right-handed neutrino signal at colliders}

Before closing the section, we would like to comment on the production and discovery potential of the two right-handed neutrinos $\nu_{1,2R}$ which govern type-I seesaw neutrino mass generation, apart from the dark matter $\nu_{3R}$ that is separately studied in the next section. 

Firstly, the neutrinos $\nu_{1,2R}$ can be produced at high energy colliders through the processes mediated by the SM bosons $W,Z$, and $H$ (see, e.g., \cite{Antusch:2016ejd}), and the new bosons $Z',H'$ (see, e.g., \cite{Das:2017deo,Das:2018tbd}). However, since $\nu_{1,2R}$ are a SM-singlet possessing only a $h^\nu$ coupling to leptons, contribution to the production rate in case of $W,Z,H$, and $H'$ exchanges is only through mixing effects, hence being very suppressed by the small mixing parameters, ${\varphi}\sim\mathcal{O}(10^{-3})$, ${\xi}\sim\mathcal{O}(10^{-2})$, $\epsilon\sim\mathcal{O}(10^{-6})$, as well as the small coupling $h^\nu\sim 10^{-5}$. That said, the neutrinos $\nu_{1,2R}$ are dominantly produced in pair due to the $Z'$ exchange, say $\sigma(pp\to Z'\to NN)$ at LHC and $\sigma(e^+e^-\to Z'\to NN)$ at ILC for $N=\nu_{1,2R}$, because both $\nu_{1,2R}$ and SM fermions have non-zero new gauge charges that set the $Z'$ force. Although the production rate in this case is suppressed by the $Z'$ mass at TeV, it opens up a possibility of discovering these heavy neutrinos in the high-luminosity experiments \cite{Das:2022rbl}. 

Next, once produced these heavy neutrinos can decay into the SM particles such as $l^\pm W^\mp, \nu_{L} Z$, and $\nu_{L} H$ through their mixings with SM neutrinos. Because of Majorana nature of $N$, the characteristic signal in experiment might be a same-sign dilepton or a trilepton plus jets (associated with quarks) and/or missing energy (associated with neutrinos) in final state. For instance, $\sigma(pp\to Z'\to NN \to l^\pm\l^\pm W^\mp W^\mp)$ is either followed by both $W$'s decay to quarks $W\to ud$ or followed by a $W$ decay to quarks $W\to ud$ and another $W$ decay to leptons $W \to l \nu_L$. Alternatively, because the $N$-$\nu_L$ mixing is very small, one of two heavy neutrinos can be long-lived, enough to be explored by the displaced vertex search experiments \cite{Das:2019fee}. A detailed study of all such processes is interesting but out of the scope of this work, which would be published elsewhere.

\section{\label{DM}Majorana dark matter}

In this section, we will show that the third right-handed neutrino $\nu_{3R}$ is a viable dark matter candidate based on the dark matter relic density and direct dark matter detection experiments. In our scenario, the dark matter candidate $\nu_{3R}$ is stable and impossibly decaying to ordinary particles because of the $\mathbb{Z}_2$ conservation. The idea of a right-handed neutrino dark matter is old, but in our model it takes a new life since it communicates with normal matter via $Z',H'$ portals, the fields of family-dependent abelian dynamics. Additionally, since the $Z$-$Z'$ and $H$-$H'$ mixings are very small, given that $v\ll\La$, we can omit extra contributions of ordinary particles due to the mixings. Also, since $H'$ only couples to ordinary particles through a mixing with $H$, the $\nu_{3R}$ candidate interacts with SM particles dominantly through the new gauge boson portal $Z'$. 

\subsection{Dark matter scattering off nuclei}

Direct dark matter detection experiments are mostly aimed at observing elastic scatters of dark matter particles on nucleons of heavy nuclei in a large detector. Furthermore, this scattering can be described, at the microscopic level, starting from effective interactions between the dark matter particle and the SM quarks. In our scenario, such interactions are dominantly contributed by the $t$-channel exchange of the new gauge boson $Z'$, as described by the following effective Lagrangian \cite{Barger:2008qd}
\be \mathcal{L}_{\mathrm{eff}} = \frac{g^2}{4c^2_W} \frac{g_A^{Z'}(\nu_R)}{m^2_{Z'}} \bar{\nu}_{3R}\gamma_\mu\gamma_5\nu_{3R}\bar{q}\gamma^\mu[g_V^{Z'}(q)-g_A^{Z'}(q)\gamma_5]q,  \ee
where the vector and axial-vector couplings of $Z'$ to quark $q$, say $g_{V,A}^{Z'}(q)$, can be directly obtained from Table \ref{tab3}, while the axial-vector coupling of $Z'$ to $\nu_{3R}$ is $g_A^{Z'}(\nu_R)\equiv C_{\nu_R}^{Z_2}$. Note that the dark matter candidate $\nu_{3R}$ is a self-conjugate Majorana fermion, so it cannot have a vector-vector coupling in the above effective Lagrangian. From this effective Lagrangian, we obtain the spin-dependent (SD) scattering cross section of the dark matter candidate $\nu_{3R}$ in the case of the neutron target, which is given by \cite{Barger:2008qd}
\be \sigma_{\mathrm{SD}} = (2!)^2 \frac{3}{\pi}\frac{g^4}{16c^4_W} \frac{m^2_{\text{DM-N}}}{m^4_{Z'}}\left[g_A^{Z'}(\nu_R)\right]^2\left[g_A^{Z'}(u)\la_u^\mathrm{N}+g_A^{Z'}(d)(\la_d^\mathrm{N}+\la_s^\mathrm{N})\right]^2,\ee
where the factor $(2!)^2$ is due to two counts of Wick contraction for the Majorana field $\nu_{3R}$. In addition, $m_{\text{DM-N}} = m_{\nu_{3R}}m_{\mathrm{N}}/(m_{\nu_{3R}}+m_{\mathrm{N}}) $ is the reduced mass of the dark matter-nucleon system, $m_{\mathrm{N}}$ is the mass of neutron, and $\la_q^\mathrm{N}$ are the fractional quark-spin coefficients, their values for the neutron are $\la_u^\mathrm{N}=-0.42$, $\la_d^\mathrm{N}=0.85$, $\la_s^\mathrm{N}=-0.88$ \cite{Cheng:2012qr}. 

\begin{figure}[!h]
\begin{center}
	\includegraphics[scale=0.4]{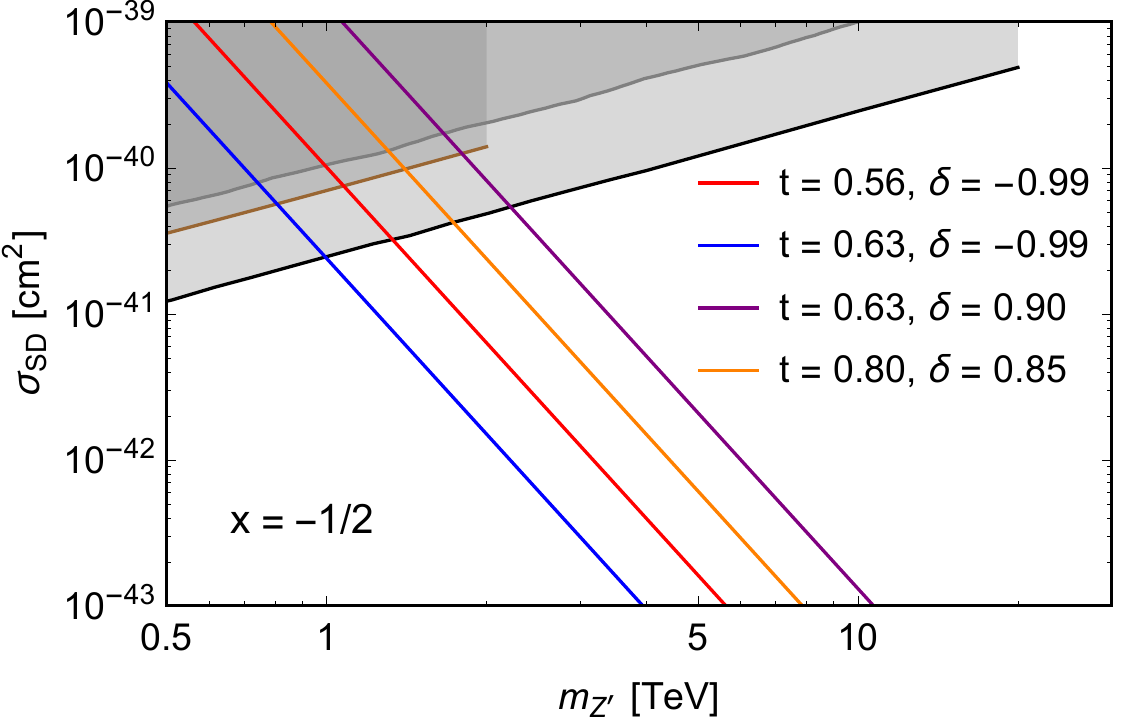}
	\includegraphics[scale=0.4]{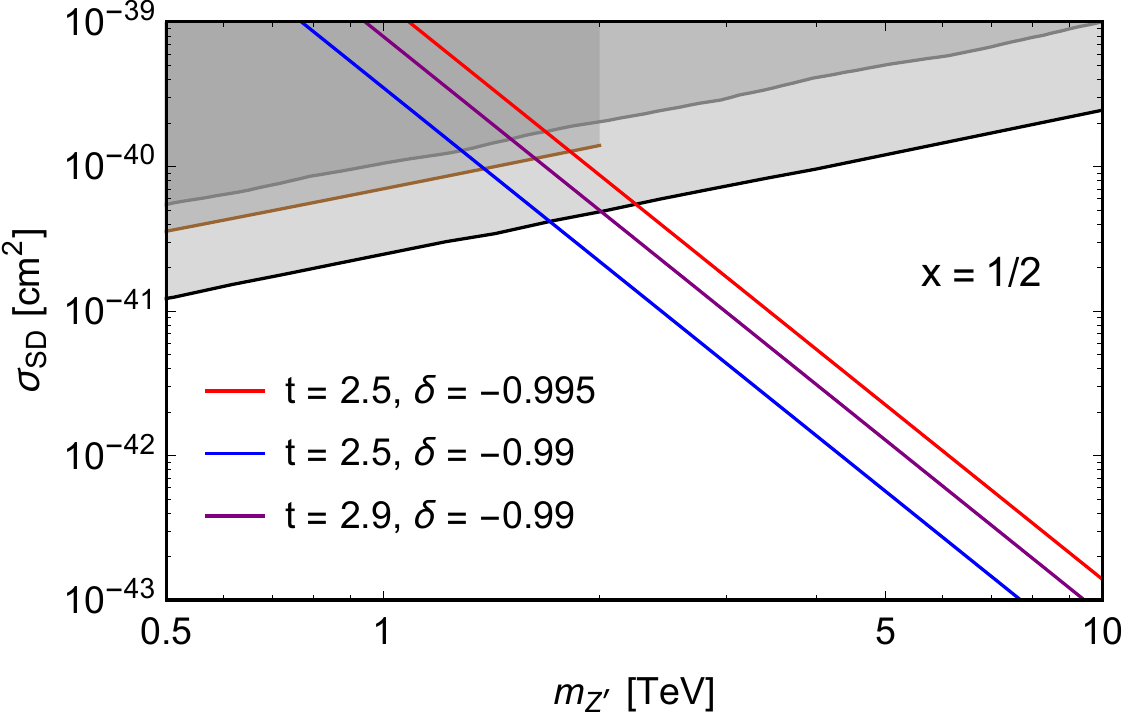}
	\includegraphics[scale=0.4]{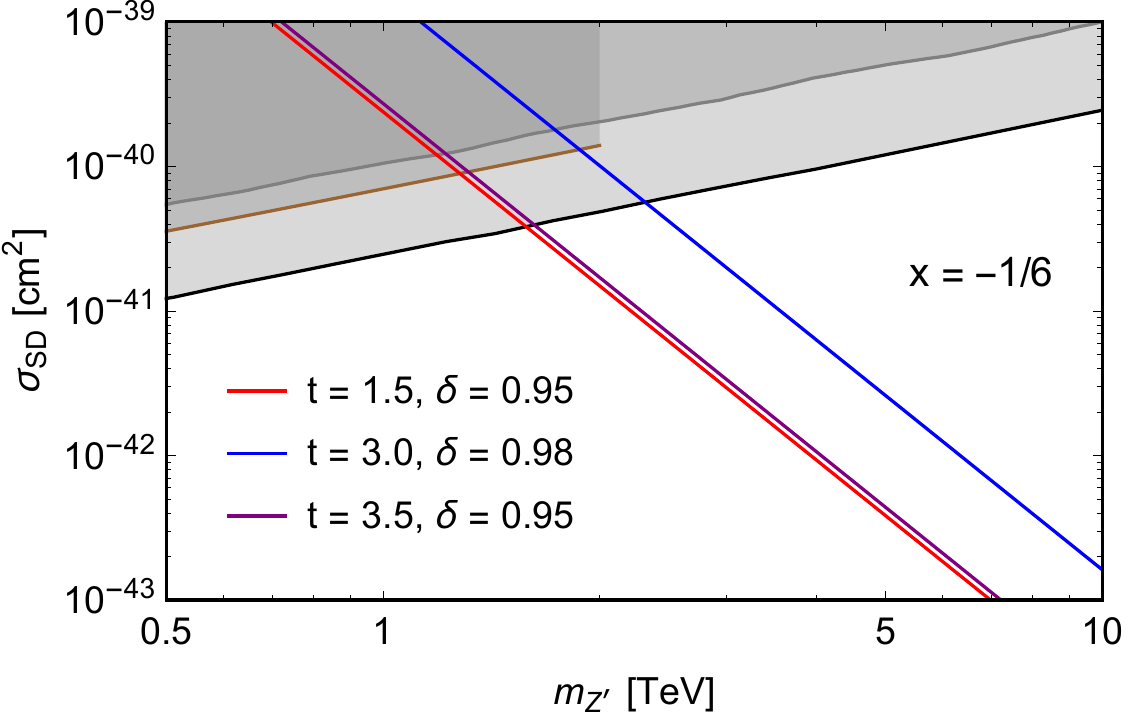}
	\includegraphics[scale=0.4]{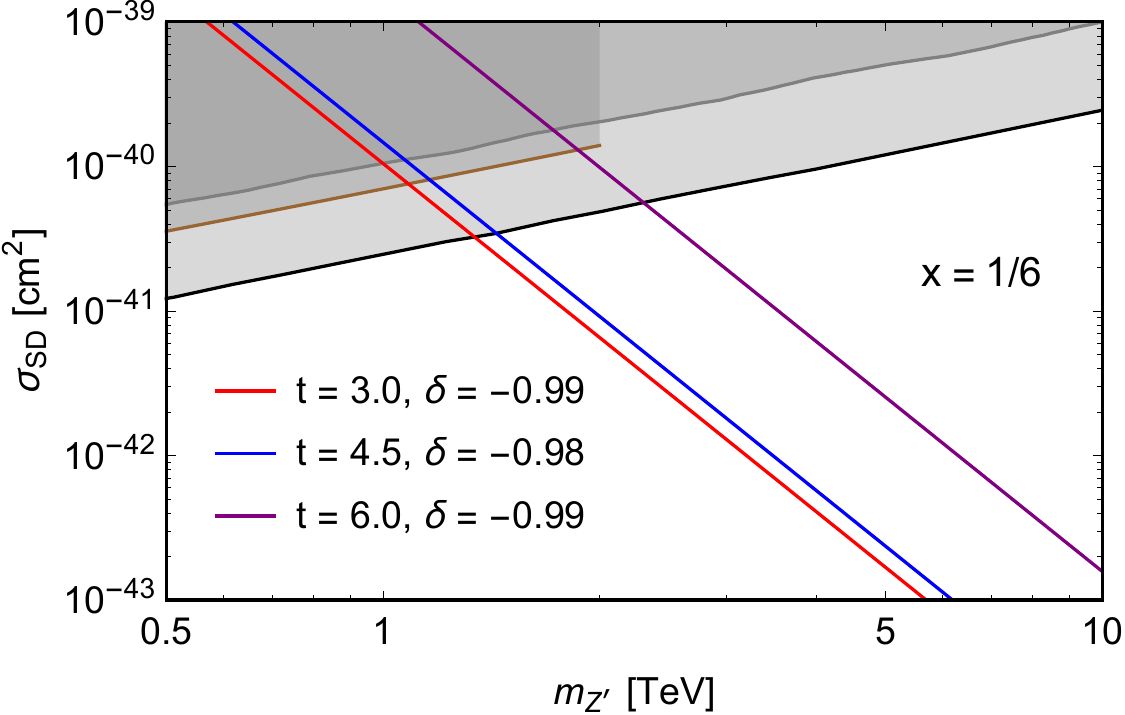}
\caption{\label{fig3}The SD $\nu_{3R}$-neutron scattering cross-section as a function of $m_{Z'}$ with $m_{\nu_{3R}}\simeq m_{Z'}/2$. The black, brown, and gray curves are the experimental results from LUX-ZEPLIN (2022) \cite{LZ:2022ufs}, XENON1T (2019) \cite{XENON:2019rxp}, and PandaX-II (2019) \cite{PandaX-II:2018woa}, respectively. The shaded regions are excluded by the these experiments. }	
	\end{center}
\end{figure}

In Fig. \ref{fig3}, we show the SD $\nu_{3R}$-neutron cross section as a function of $m_{Z'}$, according to the various values of $t$ and $\delta$ for each fixed value of $x$, assumed $m_{\nu_{3R}}\simeq m_{Z'}/2$. The selected values of $t$ and $\delta$ are the same as those in Fig. \ref{fig1}, which are within the allowed parameter spaces, as shown in Fig. \ref{fig2}.  The most stringent constraints from direct detection experiments on the SD scattering between the dark matter and the neutron target are also shown \cite{LZ:2022ufs,XENON:2019rxp,PandaX-II:2018woa}. From this figure, it is obvious that the SD $\nu_{3R}$-neutron cross section associated with the allowed parameter region obtained in the previous section is well-below the experimental limits, if the mass of new gauge boson in the TeV scale. Furthermore, the constraints from this figure are weaker than those obtained in Fig. \ref{fig1}.

\subsection{Dark matter relic abundance}

The third right-handed neutrino dark matter candidate $\nu_{3R}$ should also lead to the dark matter relic abundance consistent with the Planck collaboration \cite{Planck:2018nkj},  \be\Om_{\mathrm{DM}}h^2=0.11933\pm 0.00091.\ee 
Assuming that the dark matter candidate $\nu_{3R}$ is produced in the early universe via a process known as freeze-out, an approximate analytic solution for its relic abundance is given by
\be \Omega_{\nu_{3R}} h^2 = \frac{1.07\times 10^9 x_f}{\sqrt{g_*}M_{\mathrm{Pl}}\langle\sigma v\rangle_{\nu^C_{3R}\nu_{3R}\to\mathrm{all}}}\text{ GeV}^{-1} ,\ee
where $x_f$ is the freeze-out parameter, $g_*$ is the effective total number of degrees of freedom, $M_{\text{Pl}}$ is the Planck mass, and $\langle\sigma v\rangle$ is the thermal average of the $\nu_{3R}$ pair annihilation cross section times relative velocity. As aforementioned, the $Z$-$Z'$ mixing in the model is negligibly small as suppressed by $v^2/\Lambda^2$. In addition, we also ignore the $H$-$H'$ mixing and assume that $m_{\nu_{1,2R}}<m_{\nu_{3R}}<m_{Z'}<m_{H'}$ for simplicity. Hence, the relic density of dark matter candidate $\nu_{3R}$ is primarily dictated by its annihilations to the SM fermions and $\nu_{1,2R}$, which are all imperatively $s$-channel processes mediated by the new gauge boson $Z'$, and thus  
\bea \langle\sigma v\rangle_{\nu^C_{3R}\nu_{3R}\to\mathrm{all}} &\simeq& \fr{g^4(C^{Z_2}_{\nu_R})^2m^2_{\nu_{3R}}}{64\pi c^4_W(4m^2_{\nu_{3R}}-m^2_{Z'})^2} \left(\sum_f N_C(f)\{[g_V^{Z'}(f)]^2+[g_A^{Z'}(f)]^2\}+\frac{(C^{Z_2}_{\nu_L})^2}{2}\right)\crn
&&+\frac{g^4(C^{Z_2}_{\nu_R})^4m^2_{\nu_{3R}}}{128\pi c^4_W(4m^2_{\nu_{3R}}-m^2_{Z'})^2}\sum_{i=1,2}\left(1-\fr{m^2_{\nu_{iR}}}{m^2_{\nu_{3R}}}\right)^{1/2}\Theta\left(m_{\nu_{3R}}-m_{\nu_{iR}}\right), \eea
where $f$ represents the SM charged fermions.

\begin{figure}[h]
\begin{center}
	\includegraphics[scale=0.4]{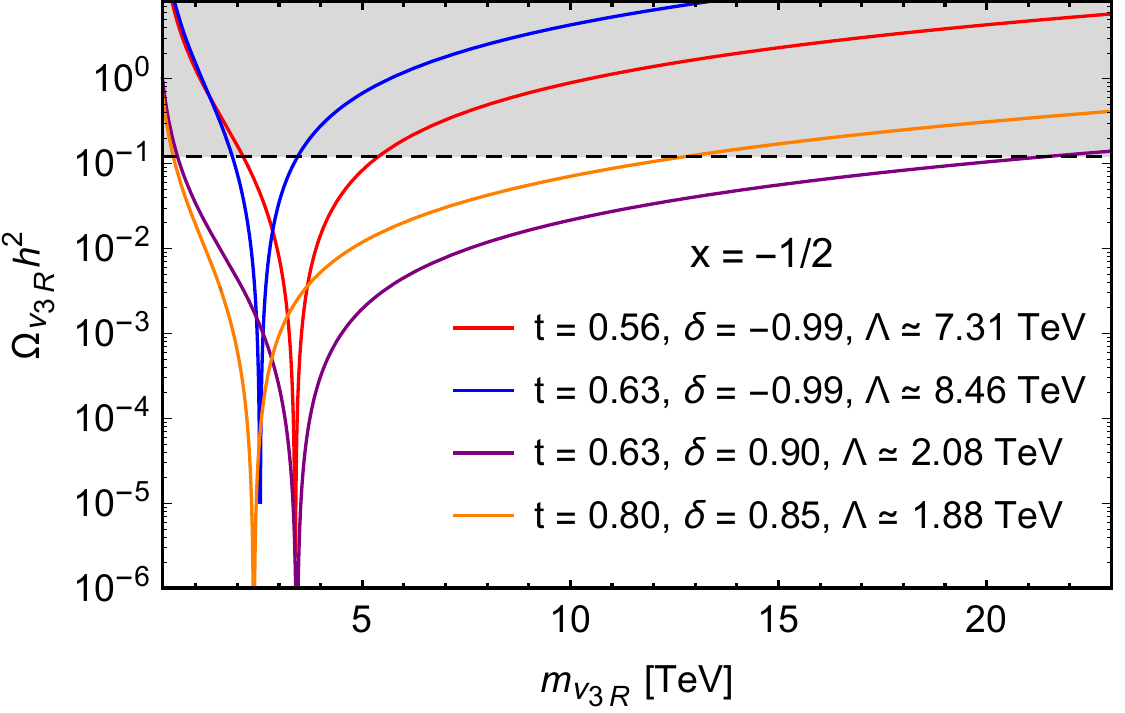}
	\includegraphics[scale=0.4]{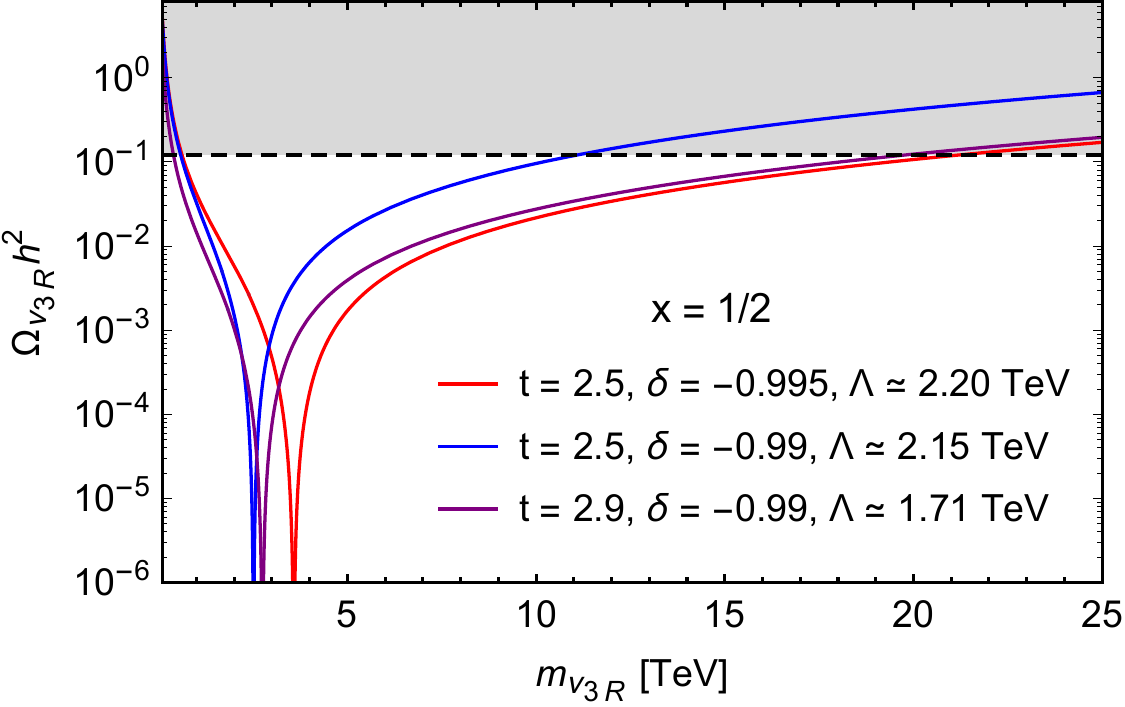}
	\includegraphics[scale=0.4]{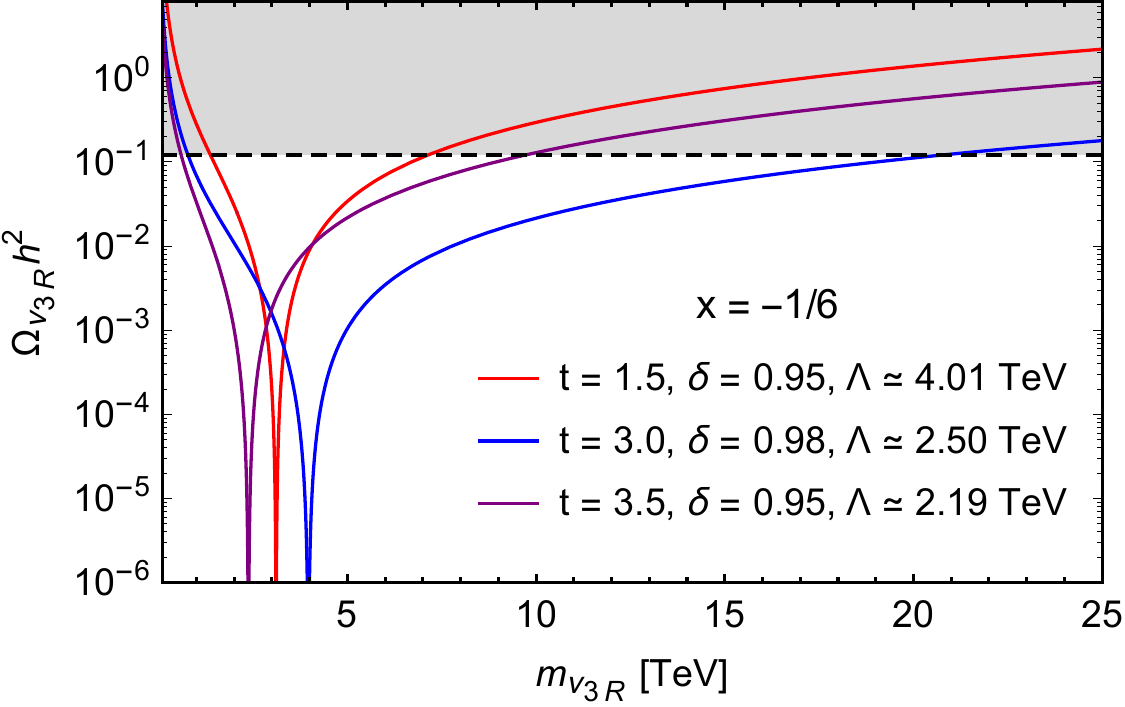}
	\includegraphics[scale=0.4]{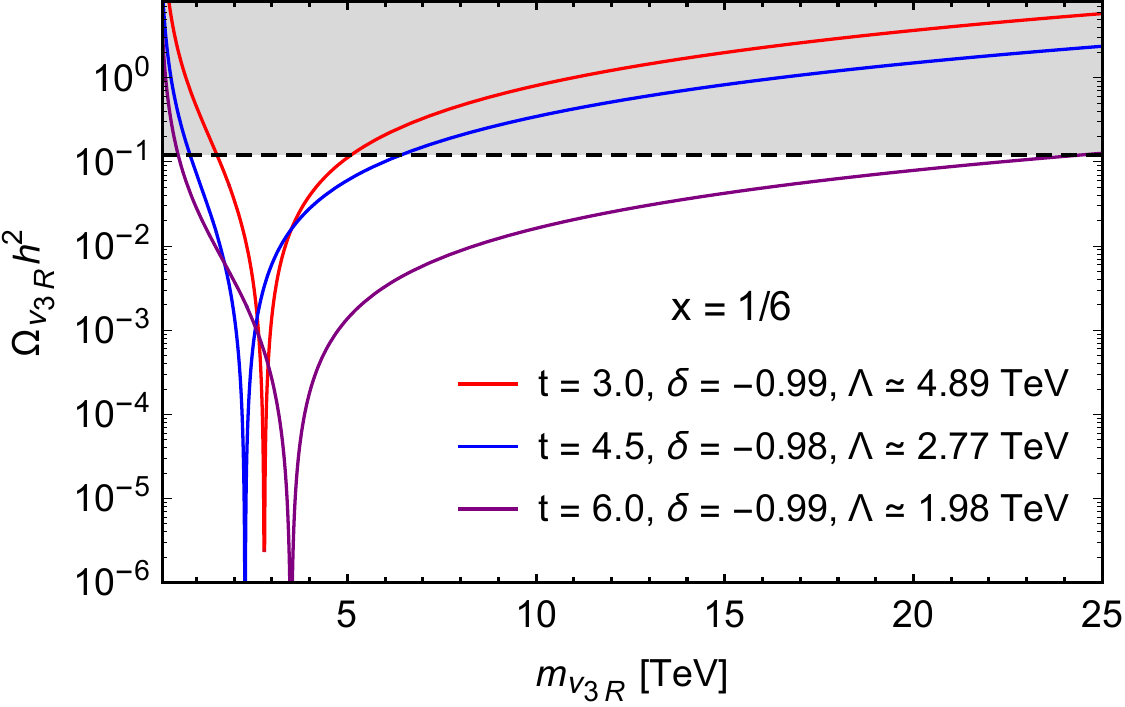}
\caption{\label{fig4}For each value of $x$, the right-handed neutrino dark-matter relic density plotted as a function of its mass for different choices of $t$, $\delta$, and $\La$, which are within the allowed parameter spaces shown in the Tab. \ref{tab4}. The shaded regions are excluded by the Planck collaboration \cite{Planck:2018nkj}.}	
	\end{center}
\end{figure}

Taking $x_f=25$, $g_*=106.75$, $M_{\mathrm{Pl}}=1.22\times 10^{19}$ GeV, and $m_{\nu_{1R}}=m_{\nu_{2R}}= m_{Z'}/3$, in Fig. \ref{fig4} we show the prediction of the model for the dark matter relic abundance as a function of the dark matter mass corresponding to various values of $t$, $\delta$, and $\La$, and for each value of $x$. The selected $t$ and $\delta$ are the same as those in Fig. \ref{fig1}, while the selected $\La$ values satisfy the CDF $W$-boson mass deviation, as in Eq. (\ref{Wmass}). From this figure, we derive the viable regions for dark matter mass as collected in Table \ref{tab5}, where the upper limits are extracted directly from Fig. \ref{fig4} while the lower limits are estimated as $m_{Z'}/2$, which simultaneously satisfy the assumptions and the constraints obtained in Sect. \ref{constraint}.

\begin{table}[h!]
\bc
\begin{tabular}{cc|cc}
\hline\hline
$\left(x,t,\delta,\Lambda \text{ [TeV]}\right)$ & $m_{\nu_{3R}}$ [TeV]& $\left(x,t,\delta,\Lambda \text{ [TeV]}\right)$ & $m_{\nu_{3R}}$ [TeV]\\  \hline
 $\left(-1/2,0.56,-0.99,7.31\right)$ & $3.39 \lesssim m_{\nu_{3R}} \lesssim 5.39$& $\left(-1/2,0.63,-0.99,8.46\right)$ & $2.54 \lesssim m_{\nu_{3R}} \lesssim 3.45$\\
 \hline
 $\left(-1/2,0.63,0.90,2.08\right)$ & $3.43 \lesssim m_{\nu_{3R}} \lesssim 21.47$& $\left(-1/2,0.80,0.85,1.88\right)$ & $2.40 \lesssim m_{\nu_{3R}} \lesssim 12.78$\\
 \hline
 $\left(1/2,2.5,-0.995,2.20\right)$ & $3.58 \lesssim m_{\nu_{3R}} \lesssim 21.22$& $\left(1/2,2.5,-0.99,2.15\right)$ & $2.51 \lesssim m_{\nu_{3R}} \lesssim 11.10$\\
 \hline
 $\left(1/2,2.9,-0.99,1.71\right)$ & $2.75 \lesssim m_{\nu_{3R}} \lesssim 19.80$& $\left(-1/6,1.5,0.95,4.01\right)$ & $3.11 \lesssim m_{\nu_{3R}} \lesssim 7.14$\\
 \hline
 $\left(-1/6,3.0,0.98,2.50\right)$ & $3.97 \lesssim m_{\nu_{3R}} \lesssim 20.79$& $\left(-1/6,3.5,0.95,2.19\right)$ & $2.38 \lesssim m_{\nu_{3R}} \lesssim 9.73$\\
 \hline
 $\left(1/6,3.0,-0.99,4.89\right)$ & $2.80 \lesssim m_{\nu_{3R}} \lesssim 5.10$& $\left(1/6,4.5,-0.98,2.77\right)$ & $2.29 \lesssim m_{\nu_{3R}} \lesssim 6.42$\\
 \hline
 $\left(1/6,6.0,-0.99,1.98\right)$ & $3.51 \lesssim m_{\nu_{3R}} \lesssim 24.46$& No data & No data\\
\hline\hline
\end{tabular}
\caption[]{\label{tab5}Viable regions of dark matter mass $m_{\nu_{3R}}$ for different choices of $t$, $\delta$, and $\La$, which simultaneously satisfy the constrains obtained in Sect. \ref{constraint}.}  
\ec
\end{table}

\section{\label{conclusion}Conclusion}
In this work, we have considered a simple extension of the SM with minimal particle content, based on the symmetry $SU(3)_C\otimes SU(2)_L\otimes U(1)_X\otimes U(1)_N\otimes \mathbb{Z}_2$, inspired by the observed fermion family number and the dark matter stability. In addition to the SM particles, the model contains only a scalar singlet and three right-handed neutrinos as the new fields. We have shown that three right-handed neutrinos are just enough to explain both the smallness of active neutrino masses and the dark matter observables. The phenomenology of the new gauge boson has been investigated through the mixing of it with the SM $Z$-like boson. We have obtained the parameter spaces that simultaneously satisfy the recent CDF result for $W$-boson mass, electroweak precision measurements, particle colliders, as well as dark matter observables, if the kinetic mixing parameter between the $U(1)_{X,N}$ gauge bosons is not necessarily small. In contrast to \cite{VanDong:2022cin} that requires at least two Higgs doublets, this work shows that only the standard model Higgs doublet presented is sufficient for recovering all phenomenological aspects when the kinetic mixing effect is included.   

\bibliographystyle{JHEP}

\bibliography{combine}

\end{document}